\documentclass[manuscript,screen]{acmart}
\usepackage{booktabs}
\usepackage{tabularx}
\usepackage{array}
\usepackage{multirow}
\usepackage{subcaption}
\usepackage{placeins}
\usepackage{enumitem}
\usepackage{xcolor}

\newcommand{\zekun}[1]{{\color{black}#1}}

\AtBeginDocument{%
  }

\setcopyright{acmlicensed}
\acmConference[Conference acronym 'XX]{Make sure to enter the correct
  conference title from your rights confirmation email}{June 03--05,
  2018}{Woodstock, NY}
\acmISBN{978-1-4503-XXXX-X/2018/06}

\begin{document}

\title{From Review to Reuse: How Post-Task Workflow Can Support Human-AI Agent Interaction}

\author{Zekun Wu}
\email{wuzekun@cs.uni-saarlaned.de}
\orcid{0000-0002-5233-2352}
\affiliation{%
  \institution{Saarland University, Saarland Informatics Campus}
  \streetaddress{Campus, 66123 Saarbrücken}
  \city{Saarbrücken}
  \state{Saarland}
  \country{Germany}
  \postcode{66123}
}

\author{Xinru Wang}
\email{xinru.wang@smart.mit.edu}
\affiliation{%
  \institution{Singapore-MIT Alliance for Research
and Technology}
  \country{Singapore}
}

\author{Rock Yuren Pang}
\email{rockpang@microsoft.com}
\affiliation{%
  \institution{Microsoft}
  \city{Redmond}
    \state{Washington}
  \country{United States}
}

\author{Chenglong Wang}
\email{chenwang@microsoft.com}
\affiliation{%
  \institution{Microsoft Research}
  \city{Redmond}
    \state{Washington}
  \country{United States}
}  
\author{Anna Maria Feit}
\email{feit@cs.uni-saarland.de}
\orcid{0000-0003-4168-6099}
\affiliation{%
  \institution{Saarland University, Saarland Informatics Campus}
  \streetaddress{E1 7, 66123 Saarbrücken}
  \city{Saarbrücken}
  \country{Germany}}

\renewcommand{\shortauthors}{Wu et al.}

\begin{abstract}
AI agents can automate tasks by turning a single natural-language request into a multi-step process spanning tools, files, and applications. Users are often left to judge that process from fragmented execution information and the final output. To make the completed process easier to understand, validate, and reuse, we investigate post-task workflows: editable, graph-based representations of an agent’s completed execution. We first analyzed 10,803 public workflow templates from n8n to characterize real-world automation practice, then developed Trace2Flow, a research probe that translates agent execution traces into interactive post-task workflows. In a study, participants (N = 20) reviewed agent executions with prompt or agent errors. We found that post-task workflows improved their understanding and error detection over a prompt-only condition, and that validation succeeded mainly when users cross-checked across multiple evidence sources. For follow-up tasks, adapting the workflow matched adapting the prior prompt in success, time, and difficulty, and was often preferred.
\end{abstract}

\ccsdesc[500]{Human-centered computing~Empirical studies in HCI}
\ccsdesc[300]{Human-centered computing~Interactive systems and tools}
\ccsdesc[100]{Computing methodologies~Intelligent agents}

\begin{teaserfigure}
  \centering
  \includegraphics[width=0.85\textwidth]{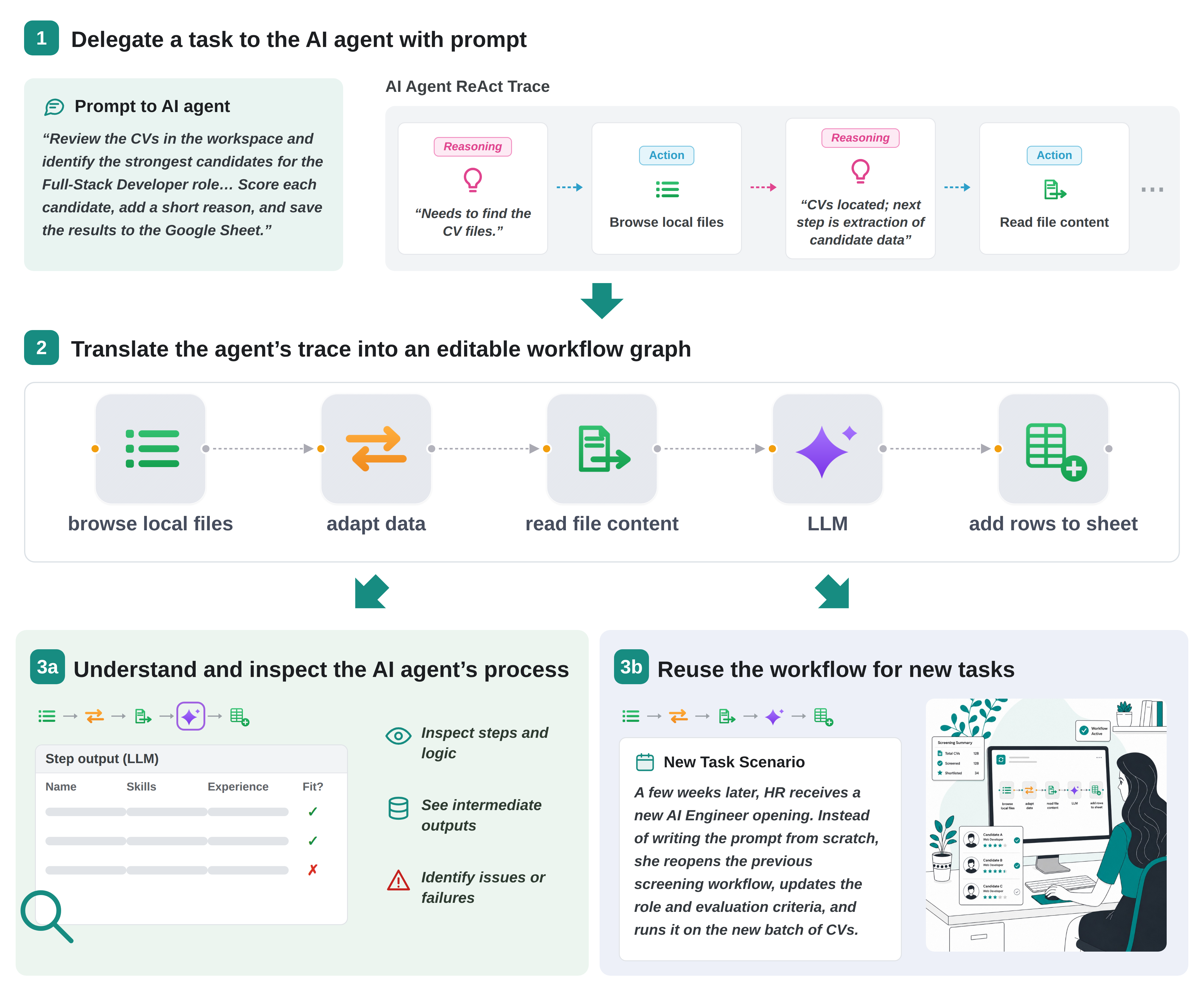}
  \caption{\zekun{Trace2Flow translates a completed prompt-based agent execution into a low-code-style workflow, allowing users to inspect how the task was carried out, validate the execution, and adapt the process for related future tasks. We use it as a research probe to explore how an editable, graph-based post-task workflow can support users in reviewing and reusing AI-agent executions. 
  }}
  \Description{Overview of Trace2Flow and the post-task workflow interaction. At the top, a user delegates a multi-step task to an AI agent through a natural-language prompt, and the agent executes the task through a sequence of reasoning and action steps. In the middle, Trace2Flow translates the completed agent execution into an editable, graph-based workflow whose connected nodes represent task operations, tool use, AI-supported operations, and their dependencies. At the bottom, the workflow supports two post-task activities: users can inspect individual steps and intermediate information to understand and validate how the task was carried out, or edit and rerun the workflow to adapt the process for a related future task.}
  \label{fig:teaser}
\end{teaserfigure}

\received{20 February 2007}
\received[revised]{12 March 2009}
\received[accepted]{5 June 2009}

\maketitle
\section{Introduction}

With the rapid development of large language models (LLMs), AI agents are increasingly being used to perform complex tasks that involve multiple rounds of reasoning, interactions with external tools, and actions across application boundaries \cite{satyanarayan2024intelligence}. The flexibility and expressiveness of natural-language prompts allow users to automate such tasks by specifying only their high-level goals. Yet, while high-level prompts make complex task delegation convenient, they leave users with the burden of reviewing how the agent interpreted their intent, verifying its actions and outputs, and identifying potential errors. Automation therefore shifts, rather than eliminates, the user’s role from carrying out the task directly to overseeing the agent’s execution, understanding what it did, and validating the result ~\cite{passi2025agentic,zhou2026should,dhanorkar2026human, faas2026}. 

Existing agent interfaces support such oversight by exposing individual execution artifacts, including action traces, tool calls, intermediate outputs, and final results \cite{mozannar2025magentic,sheng2026dills}. For example, Claude Cowork presents task progress and the agent’s current activity during execution~\cite{anthropic-cowork}. These artifacts provide visibility into an agent's behavior and can be valuable for transparency, inspection, and debugging~\cite{epperson2025interactive, barke2026agentrx}. However, understanding how a task was completed may still require users to connect information distributed across many actions, tools, and outputs. While prior work has largely focused on making individual decisions or outputs interpretable, less is known about how to represent the structure of an agent's end-to-end workflow to support human oversight, and how such a representation might benefit users when completing a complex task.

In this paper, we explore how post-task workflow representations can support users in understanding and validating agentic task execution. 
We draw inspiration from graph-based workflow representations, which externalize the operations and dependencies involved in completing a task. Such representations have a long history in low-code programming and workflow-automation tools~\cite{thuluva2020semantic, goderis2005seven}, where platforms such as n8n, Node-RED, and Zapier allow users to visually construct, configure, and execute multi-step processes across applications. In these representations, nodes typically denote operations, tools, or artifacts, while edges capture the flow of data, control, or dependencies among them. Together, these elements provide a structured, high-level view of how the different parts of a workflow fit together.

Prior work has explored using LLM agents to generate such graph-based workflows from user instructions \cite{zhou2025instructpipe}, demonstrations \cite{li2025alloy,yin2025operation}, or execution traces \cite{wu2026agenttrails,liu2026flowmind}, often formalizing the resulting graphs as specifications for subsequent execution. Although such specifications make the execution process explicit, they may constrain an agent to the operations and abstractions supported by a particular workflow system, limiting the flexibility of open-ended, prompt-based delegation. 
In contrast, we present the graph \emph{after} task completion, as a descriptive rather than prescriptive artifact that explains how the agent performed the task. This preserves the flexibility of high-level prompting while organizing the agent's actions, intermediate artifacts, and dependencies into a coherent form through which users can understand and validate the execution, diagnose errors, and reuse the validated process for related tasks.

We examine how such post-task workflow representations can support users’ interaction with AI agents in both \textit{looking backward} and \textit{looking forward}. Looking backward, a workflow may help users understand how the agent translated a high-level prompt into concrete actions, validate its decisions and outputs, and locate errors that require correction. Looking forward, the completed and validated workflow may help users reuse, adapt, or steer the process when automating related tasks. However, it remains unclear whether and how users employ workflow representations for these purposes, and when they prefer them over conventional agent outputs, execution traces, and prompting.

Concretely, we ground our study in the following research questions:
\begin{itemize}[leftmargin=*]
\item \textbf{RQ1}: How does a post-task workflow representation help users understand what the agent did, verify task outcomes,
and identify possible execution errors?
\item \textbf{RQ2}: For a related task, do users prefer to reuse and adapt a previously generated workflow, or delegate the task to an AI agent again through prompting?
\end{itemize}

We first analyzed 10,803 n8n workflow templates to understand practical workflow representations. The analysis informs the design space of workflow operations and control flow structures and also guides the construction of our study tasks. In a formative survey with 42 AI-agent users and 39 workflow-automation users we identified the difficulties users face in reviewing completed executions and the situations in which workflow representations help. Informed by these findings, we developed \textit{Trace2Flow}, a research probe that translates the AI-agent's multi-tool execution traces into an editable, executable workflow and visualizes it as a graph for users to interact with. With \textit{Trace2Flow}, 
we conducted a within-participant study with 20 regular AI-agent users to examine how access to the post-task workflow supported two activities: \textit{reviewing and validating a completed agent execution}, and \textit{completing a related follow-up task}. We found that access to the post-task workflow encouraged participants to examine the execution process rather than relying on output checking alone. This helped them recognize prompt- and agent-errors that were difficult (sometimes even impossible) to identify from the final output alone. For related follow-up tasks, we found that adapting and reusing the post-task workflow could serve as an alternative to adapting the previous prompt; in our study, the two approaches did not differ significantly in task success, completion time, or perceived difficulty. 
Nevertheless, participants tended to prefer workflow reuse when they had prior workflow-automation experience and when they only needed to make small modifications, while they prefered prompting when a previous prompt was available at all. 

Representing a completed execution as a workflow thus gives users a concrete structure for the oversight work that increasing agent autonomy leaves them with. In summary, our contributions are:

\begin{itemize}[leftmargin=*]
\item A characterization of workflow-automation practice based on 10,803 public n8n templates, describing the task domains, structural patterns, and node functions of workflows in real use, which informs both the design space of post-task workflows and the construction of our study tasks.
\item \textit{Trace2Flow}, a fully functional research probe that translates a completed prompt-based AI-agent execution trace into an interactive post-task workflow.
\item Empirical findings that post-task workflows improve users' understanding, inspection process, and validation success of a completed execution, and that workflow adaptation is a viable alternative to prompt reuse for related tasks, with preferences shaped by prior experience, adaptation scope, and prompt availability.
\end{itemize}

\section{Related Work}
\subsection{Workflow Representations for Task Automation}
\label{subsec:representation_automation}

Workflow representations have long been used to support the automation of repetitive digital tasks. Low-code workflow automation tools allow users to specify a multi-step process by connecting operations across different applications and
services, rather than implementing the entire process through conventional programming. Earlier forms of task automation, such as trigger-action programming, similarly allowed users to connect events and actions across
services to automate recurring activities
\cite{ur2014practical,ur2016trigger,zhao2021understanding}. Platforms such as n8n, Node-RED, and Zapier extend this idea through visual, editable workflows in which users can compose and configure operations provided by different applications and APIs.

More recently, AI has become integrated into workflow-automation systems, both to assist workflow construction and to perform particular operations within a workflow \cite{green2025enterprise,sharma2026agentsworkflows,zapier2026workflowindex}. Workflow automation and AI are therefore increasingly combined: AI can support flexible operations within a task, while the workflow provides an explicit and editable structure for organizing and executing the overall process. This combination makes workflow-automation practice particularly relevant to our study. We therefore examine publicly available n8n workflow templates in the next section to understand how such combined workflows are represented in practice.


\subsection{AI Agent Process Transparency}
\label{subsec:process_transparency}

As AI agents become capable of executing multi-step task autonomously, their interfaces need to support user oversight without requiring constant attention. \zekun{This creates an important interface challenge: how to make the agent’s execution process transparent to users, rather than presenting only the final result. Drawing on prior work \cite{yu2025exploring,pareek2026sensemaking}, we use process transparency to refer to how an agent reveals its plans, actions, tool use, and intermediate states \cite{mozannar2025magentic}. Such transparency has been motivated by two broad purposes: firstly, process transparency can support users’ understanding by providing a coherent view of how an agent carries out a task from start to finish \cite{mozannar2025magentic,sheng2026dills}; secondly, it can support more targeted validation of the agent’s execution and results, for example by helping users check questionable steps, intervene in the process,etc. \cite{kazemitabaar2024improving,zhou2026should,zhang2026agentracer,barke2026agentrx, wang2026xagen}.}

Previous works have proved that simply showing more execution details does not necessarily increase the agent's process transparency, on the opposite, it could makes the agent's behavior more difficult to understand \cite{sheng2026dills,zhou2026should,kazemitabaar2024improving,dhanorkar2026human}. For example, \citet{sheng2026dills} found that raw execution logs were verbose and difficult for developers to use for diagnosis. Their system, DiLLS, therefore organizes agent behavior at multiple levels, allowing developers to first see the overall plan and execution status and then check specific actions and operations when needed \cite{sheng2026dills}. Beside \textit{what} information to show, another important design question is \textit{when} users should be brought into the loop. Zhou et al. show that deciding when to ask users to check an agent involves a trade-off: asking for confirmation too often interrupts the automation and adds interaction effort, while waiting until the end can allow errors to accumulate and make recovery more costly \cite{zhou2026should}. This tension is also reflected in how people oversee agents in practice. Dhanorkar et al. found that developers did little real-time monitoring of coding agents, while post-hoc review was a more prominent form of oversight in practice \cite{dhanorkar2026human}.


Process transparency is also important for diagnosing incorrect or unexpected outcomes, because failures may arise at different points in an execution \cite{zhang2026agentracer,barke2026agentrx,epperson2025interactive}. AgentRx approaches this problem by locating the critical failure step within the execution trajectory \cite{barke2026agentrx}. Beyond locating the problem, debugging also benefits from access to the execution at that point and the ability to try alternatives. AGDebugger provides both an overview of execution history and access to individual messages for locating and testing possible corrections \cite{epperson2025interactive}.


Across this work, three recurring principles emerge for supporting process transparency in AI-agent interaction: firstly, execution information should be structured to provide an overview of the overall process while allowing users to check individual steps when needed \cite{sheng2026dills,epperson2025interactive}; secondly, the execution should remain traceable at the step level, with access to the actions and intermediate states needed to locate and investigate possible failures \cite{barke2026agentrx,zhang2026agentracer,epperson2025interactive}; thirdly, transparency and user involvement should be provided selectively, as greater visibility and more frequent intervention can improve oversight but also introduce cognitive and interaction costs \cite{zhou2026should,kazemitabaar2024improving,dhanorkar2026human}. Following these principles, we use them to guide the design of our research probe, Trace2Flow, introduced in \autoref{sec:probe}.

\subsection{Workflows for Agent Execution Review and Reuse}

Prior work has connected workflow representations with AI-agent interaction through two main routes: deriving workflows from prior agent executions and generating workflows from user-provided task specifications. Work derived from agent executions has been used both to guide future agent behavior and to support review of completed executions. Wang et al. induces reusable workflows from past agent trajectories and selectively provides them to agents to guide future task execution \cite{wang2024agent}. Liu et al. similarly build reusable workflows from successful and failed executions of related task variations, using these experiences to capture the main task steps and how the agent can detect and recover from failures \cite{liu2026reuseit}. FlowMind shows that workflow generation can be more effective when performed after task execution, using the completed execution trace to reconstruct the workflow \cite{liu2026flowmind}. Besides visualizing what the agent did, Wu et al. organizing agent trajectories into provenance graphs to help users understand how actions and intermediate data are connected across an execution \cite{wu2026agenttrails}.

A complementary line of work focuses on generating workflows directly from what users provide, such as prompts or actions\cite{zhou2025instructpipe,zhong2026chat2workflow}. For example, Zhou et al. generate editable visual workflows from natural-language instructions through three stages: node selection, pseudocode generation, and graph compilation \cite{zhou2025instructpipe}. There are also approaches generating workflow from user demonstrations rather than textual specifications \cite{li2025alloy,yin2025operation}. For example, Li et al. convert a user's step-by-step task execution into a transparent and editable workflow that can be reused and adapted for task variations \cite{li2025alloy}.

\zekun{Our work builds specifically on the first line of research, where workflow representations are derived from completed agent executions. We refer to this resulting representation as a \textit{post-task workflow}, because it is introduced after the agent has completed the delegated task. While prior work has demonstrated that such workflows can be reconstructed from execution traces, less is known about what these post-task workflows actually add for users after an agent completes a prompted task. In particular, it remains unclear whether a post-task workflow helps users understand and validate the completed execution beyond the information already available through the prompt, execution process, and result, and whether the same post-task workflow can support adaptation and reuse for a related future task.}

\section{Workflow Representations in Real-World Practice}
\label{sec:investigation}

\zekun{To ground our investigation of post-task workflows in real-world workflow-automation practice, we analyzed 10,803 publicly available n8n workflow templates collected as of July 2026. n8n \cite{n8n-workflows} is an open workflow-automation platform with a large public repository of templates created and shared by its user community. Through this investigation, we found that (1) workflows span several recurring workplace task domains, (2) they vary in size and include sequential, branching, merging, and cyclic structures, and (3) they combine task actions, data-processing, workflow-control, and AI-supported operations. We describe these findings below.}

\subsection{Task domains}
\label{subsec:task-domains}

We first examined the major task domains of the collected workflows.Although n8n provides predefined categories and tags, these labels are mainly intended to support search and discovery. They are often too broad to describe what a workflow actually accomplishes. For example, workflows for collecting recent company news before a sales call and generating sales leads from Google Maps may both be labeled “Sales,” despite involving substantially different goals, data sources, and operations. 

We therefore developed a purpose-driven classification scheme. We manually reviewed a random sample of 100 templates and grouped their primary goals into four domains: customer operations, content production, data analytics, and personal assistance, with an additional \textit{Other} category. One author of this paper first analyzed the 100 templates and came up with an initial set of labels, then the label was discussed and finalized in meetings with all authors. We then used an \texttt{gpt-5.6} to classify all 10,803 templates using this scheme (see the prompt in Appendix).




As shown in Figure~\ref{fig:n8n-task-domains}a, customer operations was the largest category, containing 3,061 workflows (28\% of the dataset). Data analytics and personal assistance each accounted for approximately 22\%, while content production accounted for 19\%; the remaining 9\% were classified as Other, such as education demo, etc.



\begin{figure*}[t]
    \centering
    \small

    \begin{minipage}[b]{0.31\textwidth}
        \centering

        \setlength{\tabcolsep}{4pt}
        \renewcommand{\arraystretch}{1.05}

        \begin{tabular}{@{}lr@{}}
            \toprule
            \textbf{Task Domain} & \textbf{$n$ (\%)} \\
            \midrule
            Customer Operation   & 3,061 (28\%) \\
            Content Production   & 2,103 (19\%) \\
            Data Analytics       & 2,336 (22\%) \\
            Personal Assistance  & 2,340 (22\%) \\
            Other                & 963 (9\%) \\
            \midrule
            \textbf{Total}       & \textbf{10,803 (100\%)} \\
            \bottomrule
        \end{tabular}

        \vspace{2pt}

        {\footnotesize
        \textbf{(a)} Primary task domains
        }

    \end{minipage}
    \hfill
    \begin{minipage}[b]{0.65\textwidth}
        \centering

        \includegraphics[
            width=0.9\linewidth
        ]{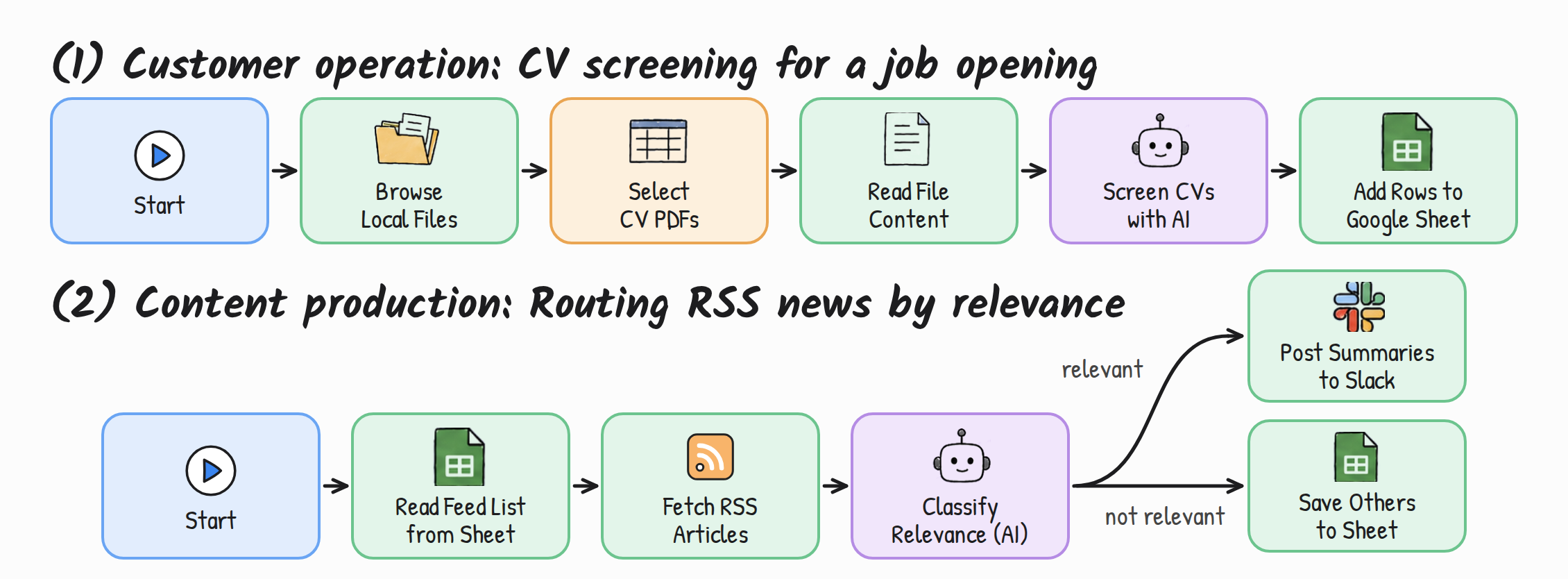}

        \vspace{2pt}

        {\footnotesize
        \textbf{(b)} Example workflows from two task domains
        }

    \end{minipage}

    \caption{
    Task domains in the 10,803 analyzed n8n workflow templates.
    (a) Distribution across primary task domains; and
    (b) representative examples from customer operations
    (CV screening) and content production (RSS-news routing).
    }
    \Description{Task domains and example workflows from the 10,803 analyzed n8n templates. Customer operations account for 28 percent of workflows, data analytics and personal assistance 22 percent each, content production 19 percent, and other tasks 9 percent. Example workflows show a multi-step CV-screening process and an RSS-news process that branches after relevance classification to different outputs.}
    \label{fig:n8n-task-domains}

\end{figure*}

\subsection{Workflow complexity}

We next examined workflow complexity from two perspectives: size and structure. Workflow size captures how extensive the represented process is, while structural complexity reflects how steps are interconnected and whether execution involves repeated paths. Specifically, we measured workflow size using the number of nodes, number of edges, and longest path (the maximum number of steps along a single execution path), and workflow structure using average node degree (the average number of connections per step) and number of cycles (paths that loop back to an earlier step).

\begin{figure*}[t]
    \centering
    \small
    \setlength{\tabcolsep}{4pt}
    \renewcommand{\arraystretch}{1.05}

    \begin{minipage}[b]{0.34\textwidth}
        \centering

        \begin{tabular}{@{}llr@{}}
            \toprule
            \textbf{Dimension} &
            \textbf{Measure} &
            \textbf{Median (IQR)} \\
            \midrule

            \multirow{3}{*}{\textit{Size}}
            & Nodes        & 17 (12--25) \\
            & Edges        & 12 (7--19) \\
            & Longest path & 8 (5--12) \\

            \midrule
            \multirow{2}{*}{\textit{Structure}}
            & Avg. node degree & 1.4 (1.1--1.6) \\
            & Cycles           & 0 (0--0) \\

            \bottomrule
        \end{tabular}

        \vspace{3pt}

        {\footnotesize
        \textbf{(a)} Workflow size and structural complexity
        }

    \end{minipage}
    \hfill
    \begin{minipage}[b]{0.62\textwidth}
        \centering

        \includegraphics[
            width=\linewidth
        ]{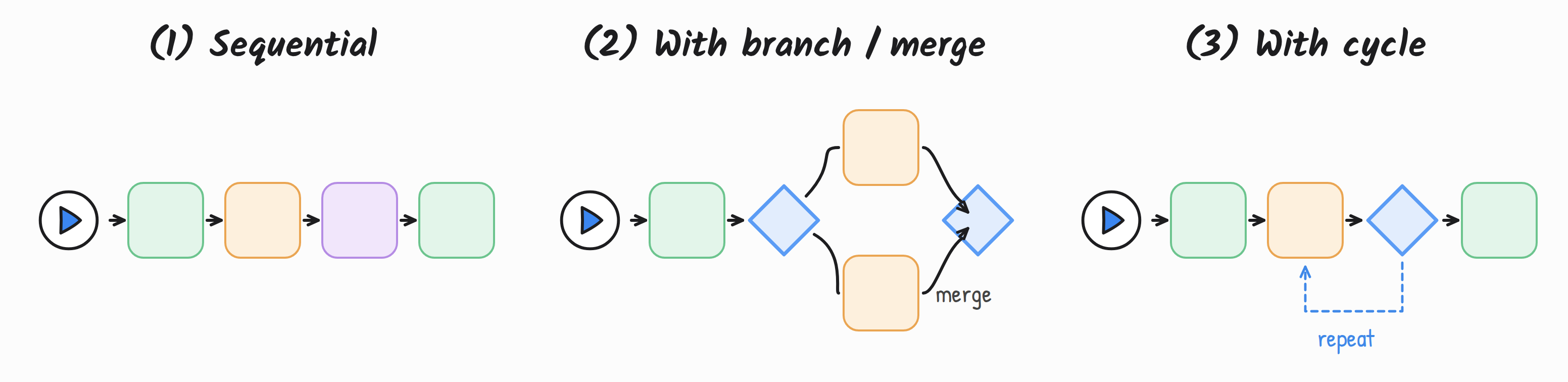}

        \vspace{3pt}

        {\footnotesize
        \textbf{(b)} Representative workflow structures
        }

    \end{minipage}

    \caption{
    Workflow complexity in the 10,803 analyzed n8n workflow templates.
    (a) Summary statistics for workflow size and structure;
    (b) representative examples of sequential, branching/merging,
    and cyclic workflow structures.
    }
    \Description{Workflow size and structural complexity in the analyzed templates. The median workflow has 17 nodes, 12 edges, a longest path of 8 steps, an average node degree of 1.4, and no cycles. Representative diagrams illustrate three structures: a sequential workflow, a workflow with branching and merging, and a workflow containing a cycle.}
    \label{fig:n8n-workflow-complexity}

\end{figure*}

As shown in Figure~\ref{fig:n8n-workflow-complexity}a, workflows contained a median of 17 nodes and 12 edges, while the median longest path contained 8 steps. Thus, although workflows could contain many operations overall, a single execution path typically traversed a smaller subset of them. The median average node degree was 1.4, indicating that workflows were generally sparsely connected, and the median number of cycles was zero. Figure~\ref{fig:n8n-workflow-complexity}b illustrates the main structural patterns represented in the dataset: purely sequential workflows, workflows containing branching or merging, and workflows containing cycles. Across the dataset, 17\% were purely sequential, 49\% contained exactly one branch or merge, and 24\% contained at least one cycle.


\subsection{Node functions}
\label{subsec:node-functions}

Then we examined the functions performed by individual nodes within each of these workflows. Again, although n8n provides official labels for its nodes, these labels primarily describe their technical category, such as Analytics, or the associated service, such as Google Sheets, rather than the functional role they play within a workflow.


Based on our review of the 195 nodes officially maintained by n8n \cite{n8npulse_nodes} and their supported operations, we distinguished four general node functions:

\begin{itemize}
    \item \textbf{Task-action:} Nodes perform substantive work through applications or services, such as retrieving data, sending messages, or updating records.

    \item \textbf{Data-processing:}  Nodes manipulate intermediate data by filtering, transforming, combining, formatting, or calculating it.

    \item \textbf{AI-supported operation:}  Nodes use an AI model or agent to perform semantic or generative work, such as extraction, classification, summarization, or generation.

    \item \textbf{Workflow-control:} Nodes determine when and how the workflow proceeds, such as starting, routing, branching, merging, waiting, or repeating execution.
\end{itemize}

\begin{figure*}[t]
\centering
\small
\setlength{\tabcolsep}{3pt}
\renewcommand{\arraystretch}{1.05}

\begin{minipage}{0.98\textwidth}
\centering
\footnotesize

\begin{tabular}{@{}
    lr
    @{\hspace{1.2em}}
    lr
    @{\hspace{1.2em}}
    lr
    @{\hspace{1.2em}}
    lr
@{}}
\toprule

\multicolumn{2}{c}{\textbf{Task-action}} &
\multicolumn{2}{c}{\textbf{Data-processing}} &
\multicolumn{2}{c}{\textbf{AI-supported operation}} &
\multicolumn{2}{c}{\textbf{Workflow-control}} \\

\multicolumn{2}{c}{\textbf{10,181 (94\%)}} &
\multicolumn{2}{c}{\textbf{8,845 (82\%)}} &
\multicolumn{2}{c}{\textbf{5,541 (51\%)}} &
\multicolumn{2}{c}{\textbf{9,384 (87\%)}} \\

\cmidrule(lr){1-2}
\cmidrule(lr){3-4}
\cmidrule(lr){5-6}
\cmidrule(lr){7-8}

HTTP Request
    & 5,428 (50\%) &
Code
    & 5,799 (54\%) &
AI Agent
    & 3,683 (34\%) &
If
    & 4,840 (45\%) \\

Google Sheets
    & 3,533 (33\%) &
Set
    & 5,637 (52\%) &
OpenAI Chat Model
    & 2,568 (24\%) &
Merge
    & 2,134 (20\%) \\

Gmail
    & 3,257(20\%) &
Split Out
    & 1,294 (12\%) &
Structured Output Parser
    & 1,772 (16\%) &
Split In Batches
    & 1,935 (18\%) \\

\bottomrule
\end{tabular}

\vspace{3pt}

{\footnotesize
\textbf{(a)} Prevalence of node-function categories and frequently used nodes
}

\end{minipage}

\vspace{8pt}

\begin{minipage}{0.96\textwidth}
\centering

\includegraphics[
    width=0.9\linewidth
]{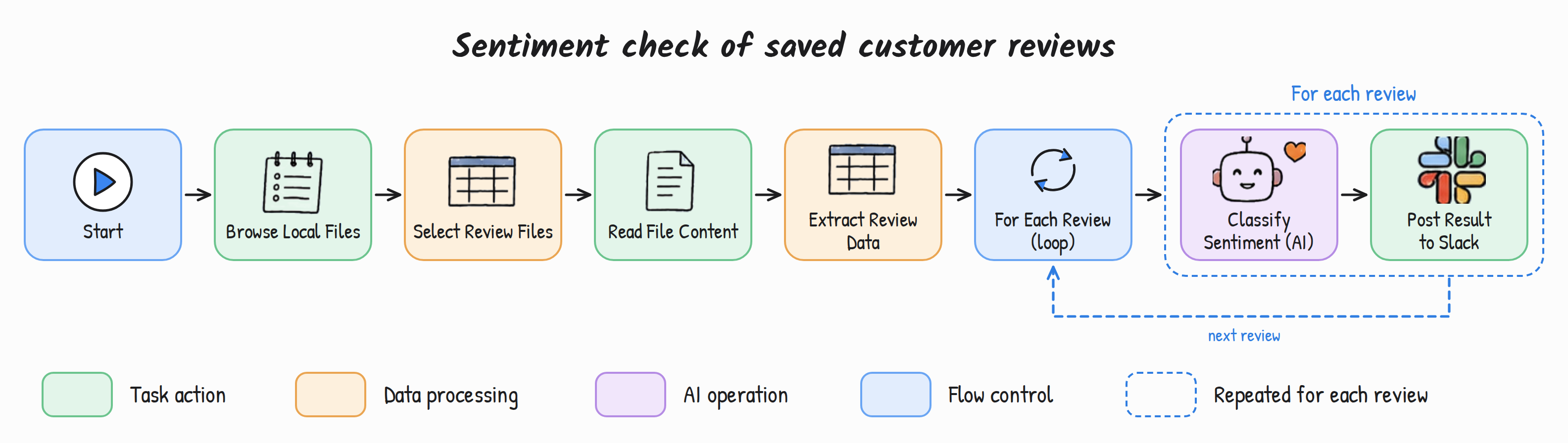}

\vspace{3pt}

{\footnotesize
\textbf{(b)} Representative workflow illustrating the functional roles of nodes
}

\end{minipage}

\caption{
Node functions in the analyzed n8n workflow templates and their roles within
a representative workflow.
(a) Prevalence of task-action, data-processing, AI-supported-operation,
and workflow-control nodes, together with frequently used nodes in each category;
(b) an illustrative customer-review workflow showing how task-action,
data-processing, AI-supported, and workflow-control operations are combined
within a multi-step process.
}
\Description{Prevalence and use of four workflow-node functions. Task-action nodes appear in 94 percent of workflows, workflow-control nodes in 87 percent, data-processing nodes in 82 percent, and AI-supported operations in 51 percent. A representative customer-review workflow combines these roles by retrieving and processing review data, iterating over reviews, classifying sentiment with AI, and posting results to Slack.}
\label{fig:n8n-node-functions}

\end{figure*}

Because a node's function can depend on how it is configured and used, we used an LLM (gpt-5.6) to classify nodes based on their type, configured operation, and workflow context across the 10,803 templates. As shown in \autoref{fig:n8n-node-functions}a, task-action nodes appeared in nearly all workflows (94\%), followed by workflow-control nodes (87\%) and data-processing nodes (82\%). AI-supported operations were also common, appearing in 51\% of workflows. What we found particularly interesting is that, besides general-purpose AI components such as AI Agent and OpenAI Chat Model nodes, many AI-supported nodes had very explicit, task-specific functions, such as information extraction, text classification, sentiment analysis, and summarization. These results show that workflow execution typically combines substantive task actions with operations that transform intermediate data, control execution flow, and incorporate AI-supported capabilities. Several individual nodes were particularly prevalent: HTTP Request (50\%) connects workflows to external APIs and services; Code (54\%) performs custom calculations or transformations; Set (52\%) restructures intermediate data for subsequent steps; and If (45\%) evaluates conditions and routes execution along different paths. Figure~\ref{fig:n8n-node-functions}b illustrates how different roles can be combined within a single process. In the illustrated customer-review workflow, task-action nodes retrieve and read review files, data-processing nodes select and extract review data, workflow-control nodes iterate over individual reviews, and an AI-supported node classifies their sentiment before the results are posted to Slack.



\section{Trace2Flow: A Research Probe for Post-Task Workflows}
\label{sec:probe}

\subsection{Formative Study and Design Goals}

To ground the design of Trace2Flow, we first drew on the general principles for supporting agent process transparency identified in \autoref{subsec:process_transparency}, and then conducted a formative study to examine whether a post-task workflow could address users’ actual needs in reviewing and reusing completed agent executions.

The formative study consisted of two complementary surveys: one with regular AI-agent users and another with users of low-code workflow-automation tools. For the AI-agent survey, we recruited participants from online communities where people discuss the use and development of AI-agent systems, including the LangChain Community, Hugging Face Forums, OpenAI Developer Community, and relevant Reddit communities, and retained 42 valid responses. For the workflow-automation survey, we recruited participants from communities and forums associated with workflow-automation tools such as n8n, Zapier, Node-RED, etc., and retained 39 valid responses. For both surveys, we included only participants who reported using the corresponding tools at least weekly. We additionally reviewed responses to the open-ended questions for relevance and interpretability, excluding submissions that were unrelated to the question or too incomplete to meaningfully interpret.

Specifically, both surveys began by asking about participants’ current practices with the corresponding tools, including which AI-agent or workflow-automation tools they used, how long they had used them, and how frequently they used them. The remaining questions focused on experiences relevant to our investigation. AI-agent users were asked to describe concrete cases in which they had difficulty reviewing a task completed by an agent, including difficulties understanding what the agent had done and determining whether the task had been completed correctly. Workflow-automation users were asked to describe cases in which an existing workflow representation helped them understand, validate, or debug a multi-step task, and how they used the workflow in those situations. Finally, both groups were asked to reflect on task reuse: when they would prefer to delegate a task through natural language and when they would instead prefer to adapt an existing workflow for a new but related task.

Based on the survey results, AI-agent users reported difficulties in understanding what an agent had done, verifying its results, and locating where errors occurred, and expressed a need for greater visibility into individual steps and intermediate information. Workflow-automation users, meanwhile, described workflow representations as useful for understanding multi-step processes, locating failures, and testing or modifying individual steps. They also highlighted workflow reuse as particularly useful when a new task was similar to an existing one, while natural-language delegation remained preferable for substantially different or exploratory tasks. These findings provided additional evidence that a post-task workflow could address needs arising in both reviewing a completed execution and adapting it for related future tasks. We report details of the formative study in the Appendix.


Based on these insights and the principles identified in prior work \cite{sheng2026dills,epperson2025interactive,barke2026agentrx,zhang2026agentracer,zhou2026should,kazemitabaar2024improving,dhanorkar2026human}, we defined three design goals for our research probe, Trace2Flow:

\textbf{DG1: Make the agent’s execution process visible and traceable.} Trace2Flow should present the AI agent’s actions, task steps, and tool calls in a single workflow graph, allowing users to trace how the final result was produced, inspect the inputs and outputs of each step, and rerun individual steps when needed.

\textbf{DG2: Support workflow-based adaptation and reuse.} When a completed task needs to be applied in a similar but new context, Trace2Flow should allow users to adapt the generated workflow by changing only the necessary inputs, steps, or tools while preserving the rest of the process.

\textbf{DG3: Complement natural-language delegation rather than replace it.} Trace2Flow should seamlessly connect prompt-based AI-agent delegation with an auto-generated editable workflow, allowing users to either describe a task in natural language or directly modify the workflow depending on the situation.

\subsection{System Design}

\subsubsection{Interface Experience}

\begin{figure*}[t]
    \centering

    \begin{subfigure}[t]{0.48\textwidth}
        \centering
        \includegraphics[width=\linewidth]{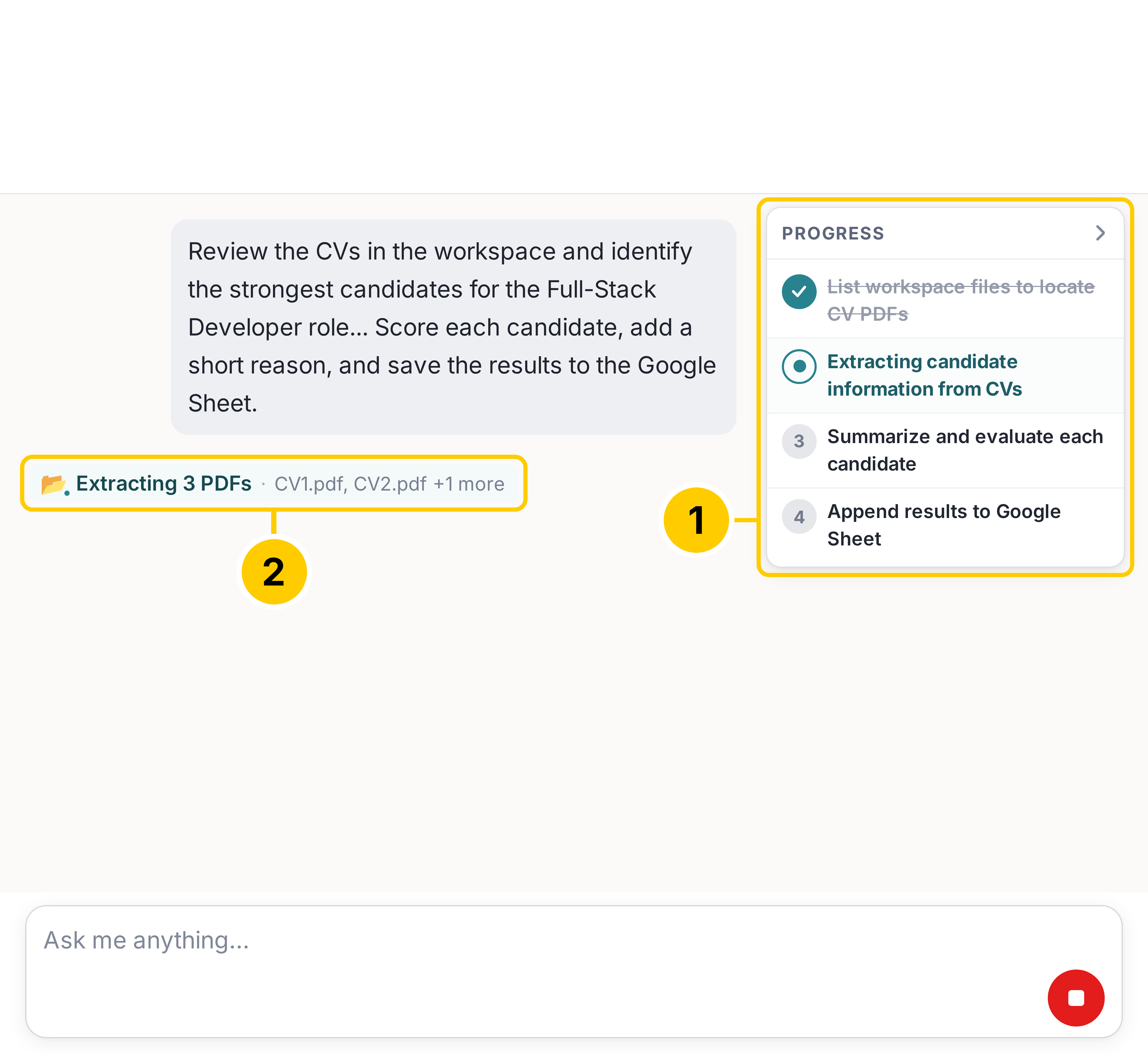}
        \caption{Task delegation and execution progress.(a) is }
        \label{fig:chat-running}
    \end{subfigure}
    \hfill
    \begin{subfigure}[t]{0.48\textwidth}
        \centering
        \includegraphics[width=\linewidth]{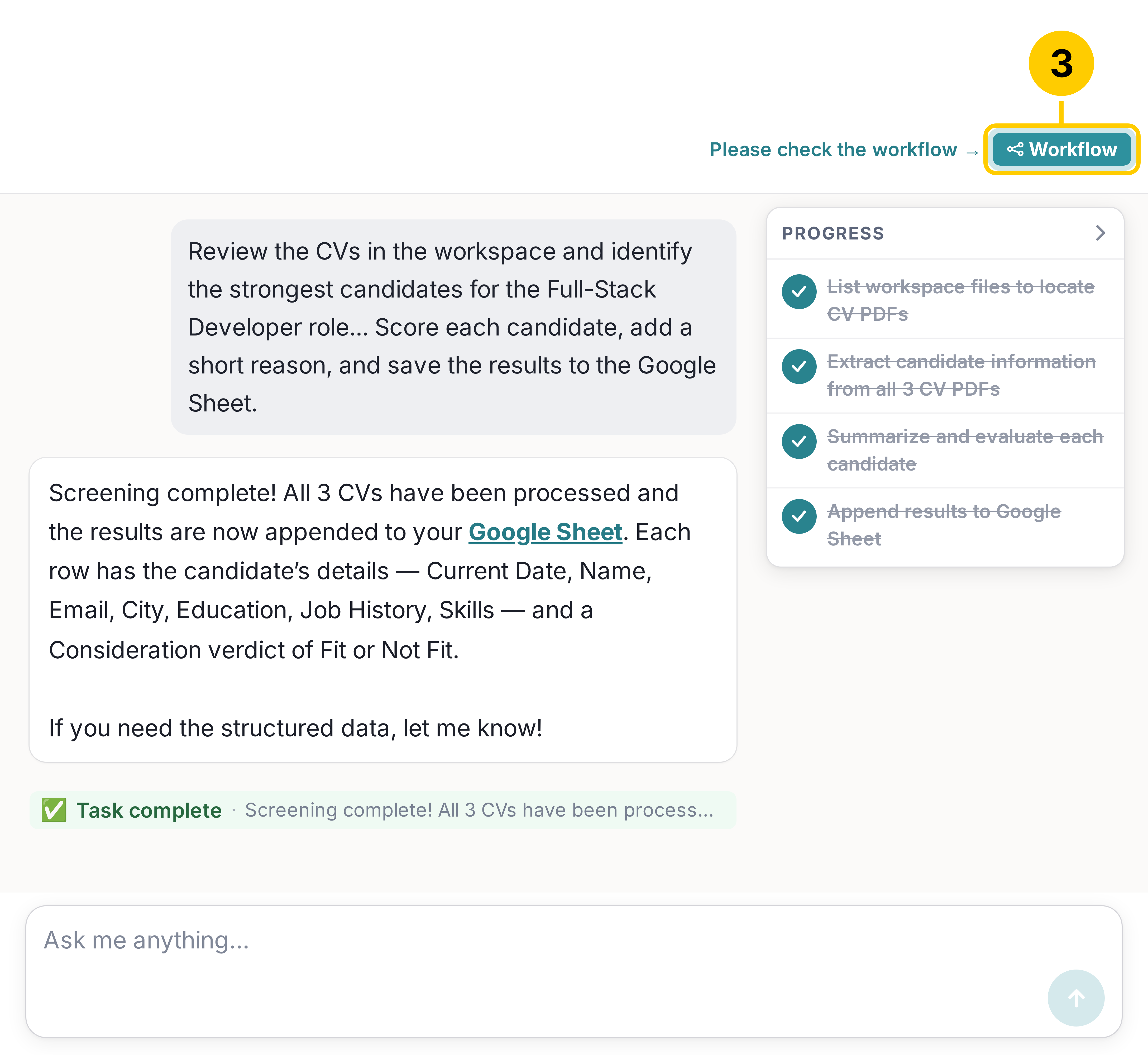}
        \caption{Task completion and execution results.}
        \label{fig:chat-complete}
    \end{subfigure}

    \par\medskip

    \begin{subfigure}[t]{0.98\textwidth}
        \centering
        \includegraphics[width=\linewidth]{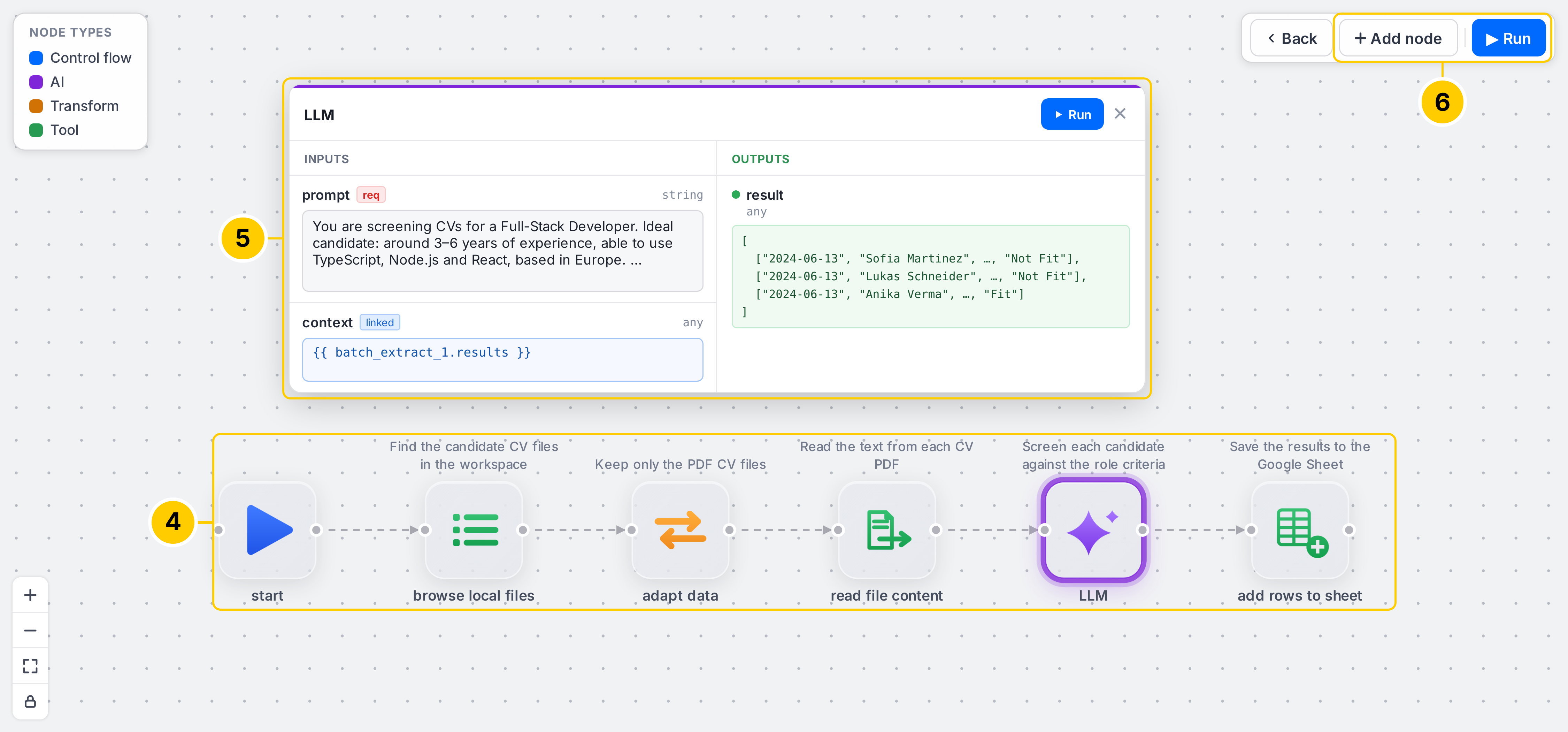}
        \caption{Generated workflow inspection and editing.}
        \label{fig:workflow-interface}
    \end{subfigure}

\caption{Trace2Flow's two primary interfaces and the transition between them. (a) During task execution, the chat interface shows the agent's progress through a step checklist (1) and surfaces its current activity (2). (b) After the task is completed, the interface presents the execution result in chat response and provides a \textit{Workflow} entry (3) for opening the translated workflow. (c) The workflow canvas represents the completed execution as a graph of connected task steps (4); selecting a node opens its detail panel (5), where users can inspect and modify the node's inputs, configuration, and outputs.Users can also make structural changes to the workflow, add new nodes, and run the full workflow to test the adapted process (6).} \label{fig:system-interface}

\Description{Trace2Flow interface across task execution, completion, and workflow review. During execution, the chat interface shows a checklist of major task steps and the agent's current activity. After completion, it shows the final result and a button for opening the generated workflow. The workflow canvas represents the completed execution as connected task nodes; users can inspect and edit a selected node's inputs, configuration, and outputs, modify the graph structure, and rerun the workflow.}
\end{figure*}

Trace2Flow provides two connected interfaces: a chat interface for prompt-based task delegation and agent execution, and a workflow canvas for reviewing, editing, and rerunning workflows generated from completed agent executions. After an agent completes a task, users can move directly from the chat interface to its generated post-task workflow; they can also reopen an existing workflow when they want to adapt and reuse a previously executed process. \autoref{fig:system-interface} illustrates this interaction across the two interfaces.

\paragraph{Chat interface.} The chat interface (\autoref{fig:chat-running}) supports natural-language task delegation while providing lightweight visibility into the agent’s ongoing execution. To approximate the interaction patterns of contemporary chat-based AI-agent systems, we modeled its execution-status display after Claude Cowork, which surfaces progress information as a task proceeds \cite{anthropic-cowork}. A progress panel (callout 1) presents the major task steps and their execution status, while an activity indicator (callout 2) shows the operation currently being performed, such as extracting candidate information from local files. After the agent completes the task, the interface presents the final output together with the completed progress checklist. Users can then select the \textit{Workflow} button (callout 3) to translate the recorded execution trace into a post-task workflow and open it on the workflow canvas.



\paragraph{Workflow canvas.}
The workflow canvas (\autoref{fig:workflow-interface}) presents the completed agent execution as an editable graph, where nodes represent individual task operations and connections represent dependencies between them (callout 4). Users can inspect the overall execution structure or select an individual node to examine its inputs, outputs, and editable parameters in the configuration panel (callout 5). They can modify these parameters and rerun individual steps when needed. For adaptation and reuse, users can also make structural changes by adding or removing nodes and changing connections between them, and then rerun the full workflow to execute the adapted process (callout 6).



\subsubsection{Implementation: Trace-to-Workflow Translation}
\label{subsubsec:trace-to-workflow}

\begin{figure}[t]
\centering
\includegraphics[width=\linewidth]{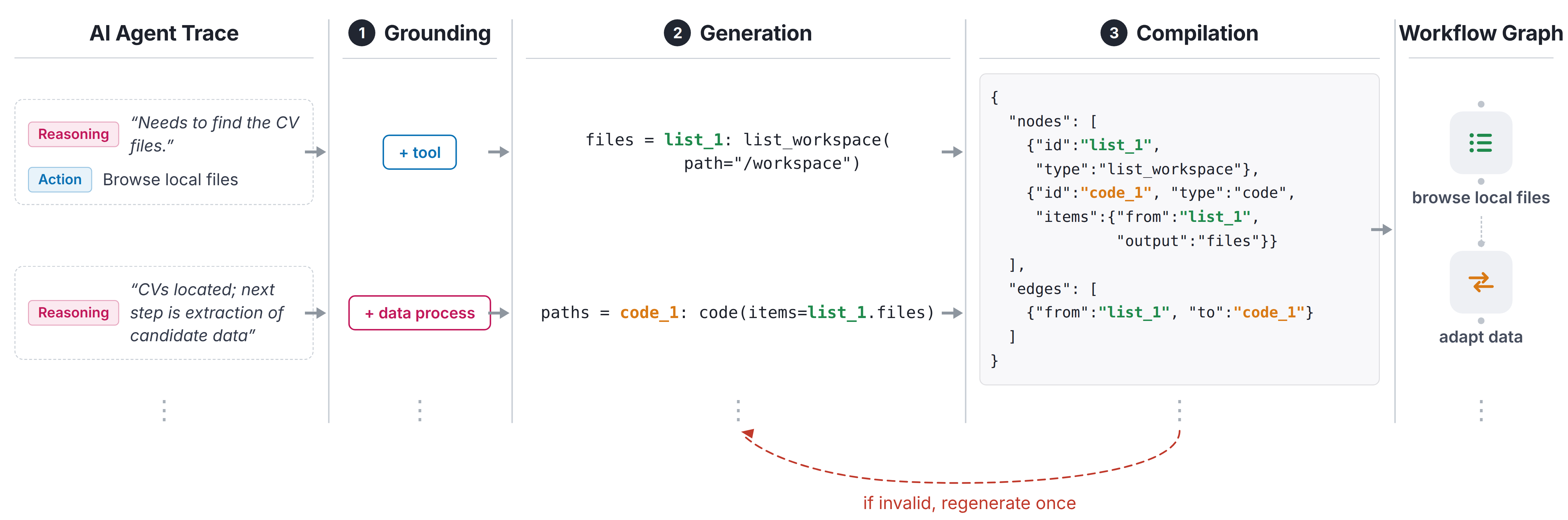}
\caption{Trace-to-workflow translation in Trace2Flow. The system first grounds candidate workflow operations in the agent's execution trace, then generates an intermediate workflow representation, and finally compiles the representation into an executable workflow graph.}
\Description{Trace-to-workflow translation pipeline. Tool actions from the AI-agent execution trace are first grounded to available workflow operations. An LLM then generates an intermediate representation containing operations and symbolic data dependencies, which is compiled into explicit workflow nodes and edges. If compilation detects an invalid representation, the error is fed back to the generation stage for one regeneration attempt.}
\label{fig:trace2flow}
\end{figure}

To support post-task validation and reuse, Trace2Flow constructs an executable workflow representation from the agent’s completed execution. We implement this component by adapting ideas from prior work on generating structured and executable workflows with LLMs ~\cite{zhou2025instructpipe}. As illustrated in \autoref{fig:trace2flow}, the translation proceeds through three stages: \textit{grounding}, \textit{generation}, and \textit{compilation}.

\paragraph{Grounding}

The grounding stage identifies which workflow operations are supported by the agent's actual execution. Trace2Flow examines the agent execution trace and matches task-relevant tool executions to corresponding workflow node specifications. For each identified operation, we retain its expected inputs and outputs together with an example output observed in the trace. This produces a task-specific node catalog that constrains the operations available to the subsequent generation stage and provides concrete evidence about the data produced by each operation. For example, as shown in \autoref{fig:trace2flow}, a workspace-listing action is grounded to a \textit{list\_workspace} node specification.

\paragraph{Generation}
\label{par:generation}

Given the grounded node catalog, the generation stage uses an LLM to reconstruct the completed execution as a compact intermediate representation (IR). The model receives the original user prompt, the agent execution trace, and the grounded node specifications. It then determines which operations are needed, how they should be configured, and how data should pass between them. 

As illustrated in \autoref{fig:trace2flow}, the IR represents both individual operations and their data dependencies. For example, the output of the workspace-listing operation is referenced as the input to a later data-processing operation. In this representation, a reference such as \textit{list\_1.files} indicates that a downstream node consumes the \textit{files} output produced by the node \textit{list\_1}. 

\paragraph{Compilation}
\label{subsubsec:compilation}

The compilation stage converts the generated IR into the structured representation required by the Trace2Flow workflow canvas. The compiler instantiates the corresponding workflow nodes, converts symbolic references into explicit input bindings and graph edges, and checks that all referenced nodes exist and that outputs are compatible with the expected inputs of downstream nodes.

For example, the dependency represented by \textit{list\_1.files} in the IR is compiled into an input binding between the two corresponding nodes and an edge connecting them in the workflow graph, as illustrated in \autoref{fig:trace2flow}. If a check fails, the system provides the LLM with the generated IR, the compiled representation, and the detected error for one additional regeneration attempt. The resulting workflow can then be executed and displayed on the workflow canvas.

Before the user study, we iteratively developed Trace2Flow using the task set described in Section 5. We applied the trace-to-workflow translation to agent execution traces, inspected the resulting workflow structure and executability, and refined the probe to address recurring translation issues. We then validated the final implementation on the same task set. Additional details on this development and validation process are provided in the Appendix.

\section{User Study}
\label{sec:user_study}

We conducted a user study to examine two roles of post-task workflows: helping users understand and validate a completed agent execution, and enabling reuse for a related future task. To ground the study in real-world workflow-automation practice, we drew on our workflow investigation in \autoref{sec:investigation} to construct five task pairs adapted from real n8n workflows. Each pair consisted of a \textit{review task} and a closely related \textit{follow-up task} that preserved most of the original process while introducing a limited change. We delegated the review tasks to an AI agent and selected executions containing either a prompt error or an agent error, creating concrete cases in which participants needed to validate whether the task had been completed correctly. We then reproduced these executions consistently across participants using a Wizard-of-Oz setup and compared how participants reviewed the execution and completed the follow-up task with or without access to the post-task workflow.

\subsection{Constructing the Study Task Set}
\label{subsec:task-delegation-assessment}

\subsubsection{Study Task Pairs}
\begin{table*}[t]
    \centering

    \small
    \begin{tabularx}{\textwidth}{@{}lXX@{}}
    
        \toprule
        \textbf{TID} & \textbf{Review Task} & \textbf{Follow-up Task} \\
        \midrule

        T0 &
        Create three groups of fictional users, save each group to a separate file, and calculate subscription statistics. &
        Combine the three groups into a summary report with overall and group-level statistics. \\
\midrule
        T1 &
        Screen three CVs for a Full-Stack Developer role and record the evaluations in Google Sheets. &
        Screen the same candidates for an AI Engineer role using different hiring criteria. \\
\midrule
        T2 &
        Collect recent RSS articles, identify those about Harness Engineering, summarize the relevant ones for Slack, and save the others to Google Sheets. &
        Repeat the task for World Model articles instead. \\
\midrule
        T3 &
        Collect list of books' detailed information from website, infer their genres, sort them by price, and save all books plus separate fiction and non-fiction lists to Google Sheets. &
        Repeat the task for another list of books, but save all results to a single sheet. \\
\midrule
        T4 &
        Process customer reviews, filter out reviews already stored in Google Sheets, classify the sentiment of the new reviews, and append them. &
        Process a new set of reviews, classify their sentiment, and append all of them without checking against existing reviews. \\

        \bottomrule
    \end{tabularx}
        \caption{Review and follow-up task pairs used in the user study.}
    \label{tab:candidate-followup-tasks}
\end{table*}

To construct the study task set, we selected five review tasks represented by real n8n workflows examined in \autoref{sec:investigation}, simplifying them where necessary to keep the tasks manageable for the study. As indicated in \autoref{tab:candidate-followup-tasks}, these tasks (T0--T4) covered three major task domains identified in \autoref{subsec:task-domains}: data analytics in T0 and T3, involving simulated user-data generation and book-information organization; customer operations in T1 and T4, involving candidate screening and customer-review processing; and content production in T2, involving news-article processing. The tasks also involved different combinations of tools and data sources, including local files, web sources, Slack, Google Sheets, etc.. We used the corresponding n8n workflows to gauge the complexity of each task, aiming for tasks that were manageable in the study but still comparable to workflows seen in practice. Specifically, the reference workflows had longest paths of five to ten steps and included different structural patterns. For example, T0 involved repeated processing through a cycle, whereas T2 used branching to route different article groups to Slack and Google Sheets. AI served a specific, explicit function within each review task: generating simulated user data in T0, evaluating candidate fit in T1, classifying and summarizing news articles in T2, inferring book genres in T3, and classifying review sentiment in T4.

For each review task, we then constructed a closely related follow-up task that retained most of the original task while changing a limited part of it. For example, T1 changed the candidate profile while retaining the screening process, whereas T3 used another set of books but changed how the results were organized in Google Sheets. This eventually produced the five task pairs summarized in \autoref{tab:candidate-followup-tasks}.

\subsubsection{Prompt Construction and Agent Execution}

Two authors independently constructed prompts for each task in \autoref{tab:candidate-followup-tasks}. They were provided with a task context taken from corresponding n8n workflow page directly describing the task goal, available inputs and tools, constraints, expected outputs, and any important process requirements.

For example, the context for T0 required the agent to generate three groups of fictional users, save each group as a JSON-formatted text file, and compute group-level subscription statistics. One resulting prompt began: \textit{``Generate three groups of ten fictional users and save the groups one after another, for example, using \texttt{file\_0.txt} for group 1, Each file should hold a JSON object ...''} For each review task, we constructed three to five prompts and used them to delegate the task to the AI agent, recording the resulting execution traces and outputs.

\subsubsection{Error cases}

\begin{table*}[t]
\centering
\small
\setlength{\tabcolsep}{4pt}
\renewcommand{\arraystretch}{1.15}

\begin{tabularx}{\textwidth}{
    @{}
    c
    >{\hsize=1.0\hsize\raggedright\arraybackslash}X
    >{\hsize=1.0\hsize\raggedright\arraybackslash}X
    @{}
}
\toprule
\textbf{TID} &
\textbf{Relevant Prompt Excerpt} &
\textbf{Selected Error} \\
\midrule

T0 &
\textit{``Generate three TXT files to simulate three groups of users and
save them one after another as local text files, using
\texttt{file\_0.txt} for group 1, for example.''} &
Agent error: the AI agent processed the first group separately but
combined the remaining two groups. \\
\midrule

T1 &
\textit{``The candidate should be based anywhere in Europe.''} &
Agent error: the AI agent incorrectly treated the fictional location in
the CV as European. \\
\midrule

T2 &
\textit{``Classify the passages as relevant to Harness Engineering or not, summarize
the latter group and post it to Slack, and push the former to Google
Sheets.''} &
Prompt error: the prompt mixed the relevant and non-relevant groups. \\
\midrule

T3 &
\textit{``Update all 20 books to Sheet 1 and the fictional ones to
Sheet 2.''} &
Prompt error: the prompt did not specify that non-fiction books should be
added to Sheet 3 or that the books should be sorted by price. \\
\midrule

T4 &
\textit{``Update the new reviews to Google Sheets.''} &
Agent error: the AI agent failed to capture the requirement to update
only the new reviews and instead pushed all reviews to Google Sheets. \\

\bottomrule
\end{tabularx}
\caption{Prompt and agent errors selected for the review tasks}
\label{tab:prompt_execution_deviations}
\end{table*}

To construct review cases in which participants had something meaningful to assess, we examined the recorded executions for two types of errors. We first checked whether each prompt accurately captured the intended task requirements described in the task context. We then checked whether the agent carried out the requirements expressed in the prompt. Based on this assessment, we selected one case for each review task: two contained a \textit{prompt error}, where requirements were omitted or incorrectly expressed (T2 and T3), and three contained an \textit{agent error}, where the agent misinterpreted or failed to follow the prompt (T0, T1, and T4). The selected cases are summarized in \autoref{tab:prompt_execution_deviations}.

\subsection{Study Design}
\label{subsec:study_design}
\begin{figure}[t]
\centering
\includegraphics[width=\linewidth]{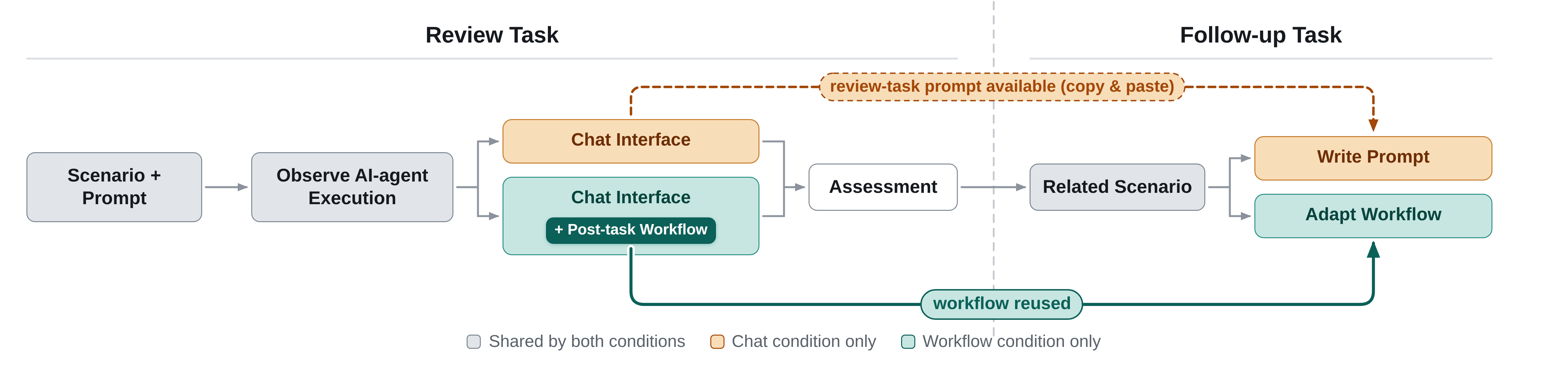}
\caption{Overview of the study task flow, from the reproduced AI-agent execution in the review task to participant completion of the related follow-up task under the \textit{chat} or \textit{workflow} condition.}
\Description{Study task flow for the two conditions. All participants receive a scenario and prompt, observe a reproduced AI-agent execution, and assess the completed task. In the chat condition, participants review the execution through chat and complete the related follow-up task by writing or adapting a prompt. In the workflow condition, participants additionally review the post-task workflow and complete the follow-up task by adapting and rerunning that workflow.}
\label{fig:task_procedure}
\end{figure}

We used a within-participant design with two conditions that differed in whether participants had access to the post-task workflow. Each task session consisted of a \textit{review task} followed by its corresponding \textit{follow-up task}, as illustrated in \autoref{fig:task_procedure}. In the \textit{chat condition}, participants reviewed the completed agent execution through the chat interface and completed the follow-up task by writing a new prompt, with the review-task prompt available for reference. In the \textit{workflow condition}, participants could additionally access the post-task workflow during review and completed the follow-up task by adapting and rerunning this workflow. We counterbalanced the condition order and task-pair assignment across participants to avoid ordering and task-specific effect.

\subsubsection{Participants Recruitment}
\label{subsubsec:recruitment}

We recruited active AI-agent users with and without prior workflow-automation experience so that we could examine how both groups understood and used the workflow representation. Participants were recruited through AI-agent and workflow-automation communities, including the LangChain Community, Hugging Face Forums, OpenAI Developer Community, relevant Reddit communities, and forums associated with n8n, Node-RED, and Zapier, as well as through our university community.

To be eligible, all participants were required to use AI-agent tools, such as ChatGPT Agent, Claude Cowork, Copilot, at least weekly. We further classified participants as having recent workflow-automation experience if they had built or edited a workflow using tools such as n8n, Node-RED, Zapier, at least once in the past month. In total, we recruited 22 participants; 20 completed the study and were included in the analysis (12 male, 8 female; ages 18--32). Of these, 7 had recent workflow-automation experience and 13 did not.
The study was approved by the institutional ethics review board. Participants were compensated with a €15 gift card.



\subsubsection{Study Procedure}
\label{subsubsec:study_procedure}
We conducted the study in both in-person and remote settings. Remote participants accessed the study application through the remote-control function in Microsoft Teams.

The study began with a tutorial using the T0 task pair to familiarize participants with the interface and study procedure. Participants then completed two or three study task sessions using task pairs drawn from T1--T4. Across the first two sessions, each participant experienced both study conditions. When time allowed, participants completed a third session. We did not require participants to complete all four task pairs to reduce fatigue and keep the overall study duration manageable.

As illustrated in \autoref{fig:task_procedure}, in each session, participants first received the workplace scenario and prompt for the review task. We then used a Wizard-of-Oz setup to reproduce the corresponding AI-agent execution using the execution trace and outputs collected in \autoref{subsec:task-delegation-assessment}. This ensured that participants assigned the same task observed the same execution and error and, in the workflow condition, the same post-task workflow.

After observing the execution, participants judged whether the task had been completed successfully and explained their reasoning. We then asked them to describe the agent's main steps, tools, and overall execution process, including how steps and data were organized. Participants also rated their overall understanding of the agent's execution on a 5-point scale.

Participants then completed the corresponding follow-up task according to their assigned condition. After the task sessions, we conducted a semi-structured interview about how the two conditions supported reviewing the agent execution and completing a related task, as well as when participants would prefer prompting or workflow reuse for similar future tasks. The full workplace scenarios, review-task questions, and interview protocol are provided in the Appendix.

\subsection{Measures and Data Analysis}
\label{subsec:data_analysis}

Our analysis was organized around the two research questions: RQ1 concerns how post-task workflows support understanding and validation of a completed agent execution, while RQ2 concerns their adaptation and reuse for related follow-up tasks. Across these two parts, we drew on participants’ task responses and ratings, screen recordings of their interactions with the probe interface, semi-structured interview transcripts, and the artifacts produced during the follow-up task, including submitted prompts in the chat condition and adapted workflows in the workflow condition.

\paragraph{RQ1.}
We examined participants’ understanding and validation of the agent’s execution using both quantitative measures and qualitative evidence. The quantitative measures were: (1) \textit{step precision and recall}, based on the agent’s major steps reported by participants; (2) \textit{tool precision and recall}, based on the tools participants reported; (3) \textit{process-comprehension accuracy}, based on predefined questions about how the task was carried out; (4) \textit{self-reported understanding}, rated on a 5-point scale; and (5) \textit{validation success}, a binary measure indicating whether participants correctly judged task completion. For the step and tool measures, we compared participants’ reported items with ground-truth lists: precision was the proportion of reported items that were correct, while recall was the proportion of ground-truth items that participants identified. Process-comprehension accuracy was calculated as the proportion of predefined process questions answered correctly.

To understand how participants arrived at these judgments, we analyzed screen recordings and interview responses. We manually coded the recordings for observable review actions, such as checking task inputs and outputs, prompts, and workflow information, and examined how participants combined these sources when forming their judgments. We also analyzed relevant interview responses to understand how participants perceived the different interface artifacts as supporting their understanding and validation of the agent’s execution.

\paragraph{RQ2.}
Similarly, we used both quantitative and qualitative analyses to answer RQ2. To examine participants’ performance and experience in the follow-up task, we measured: (1) \textit{task success}, a binary measure indicating whether participants successfully completed the follow-up task; (2) \textit{completion time}, recording how long participants took to complete the task; and (3) \textit{perceived difficulty}, rated on a 5-point scale. To understand how participants approached the follow-up task, we analyzed the prompts submitted in the chat condition, the workflows adapted in the workflow condition, screen recordings, and interview responses. We further analyzed interview responses to examine participants’ experiences and preferences for prompting or workflow reuse, including how these preferences related to prior workflow-automation experience and the scope of the required adaptation.

Across the two RQs, we used two quantitative analysis strategies according to the form and role of the measures. For the non-binary measures—\textit{step and tool precision and recall}, \textit{process-comprehension accuracy}, \textit{self-reported understanding}, \textit{completion time}, and \textit{perceived difficulty}—we aggregated observations within each participant and condition, yielding one paired value for the chat and workflow conditions, and compared them using two-sided Wilcoxon signed-rank tests. For the two binary outcomes, \textit{validation success} and \textit{follow-up task success}, we retained individual task-session observations for two reasons. First, participant-level aggregation would reduce these binary outcomes to relatively coarse values because each participant contributed only one or two sessions per condition. Second, retaining session-level observations allowed us to account for variation across tasks, which was particularly important for these direct measures of whether participants successfully validated or completed a task. We therefore analyzed the two binary outcomes separately using logistic mixed-effects models across all available sessions (N=46), with condition as the primary predictor, task and session number as covariates, and participant as a random intercept.Both models followed the specification: $\text{Success} \sim \text{Condition} + \text{TID} + \text{Session Number} + (1 \mid \text{Participant ID})$.

\section{Results}
Corresponding to our two RQs, we report the results in two parts. We first examine participants’ understanding and validation of the completed agent execution, including their review actions and patterns. We then examine their performance and experience in the follow-up task, including task success, completion time, perceived difficulty, and preferences for prompting versus workflow reuse.

\subsection{Reviewing Agent Execution: Understanding and Validation}

\subsubsection{Understanding the Agent’s Execution Process}

\begin{figure*}[t]
    \centering

    \begin{subfigure}[t]{0.23\textwidth}
    \centering
    \captionsetup{skip=3pt}
    \includegraphics[
        width=\linewidth,
        trim=0 20 0 0,
        clip
    ]{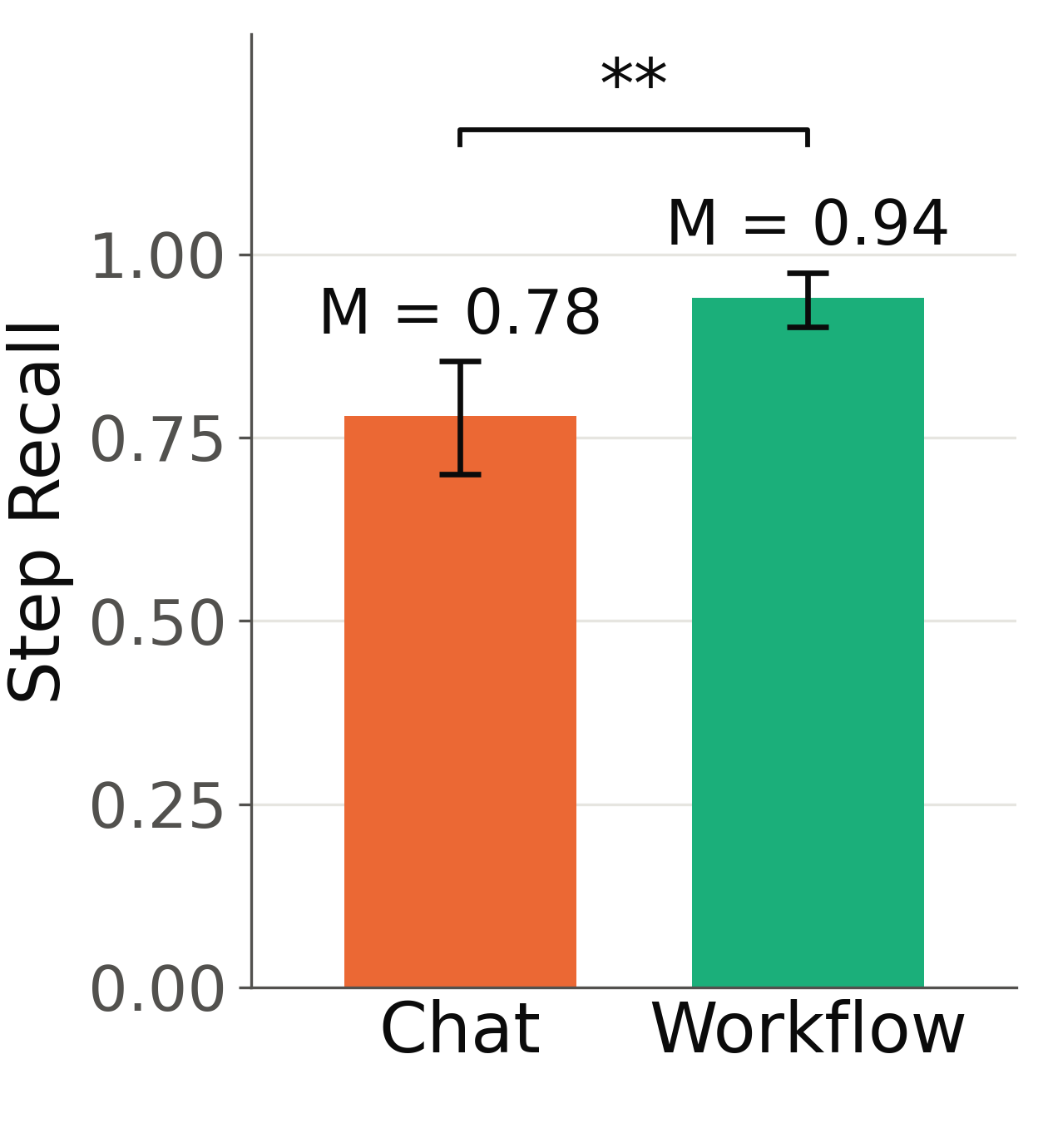}
        \caption{Step Recall}
        \label{fig:step_recall}
    \end{subfigure}
    \hfill
    \begin{subfigure}[t]{0.23\textwidth}
        \centering
    \captionsetup{skip=3pt}
    \includegraphics[
        width=\linewidth,
        trim=0 20 0 0,
        clip
    ]{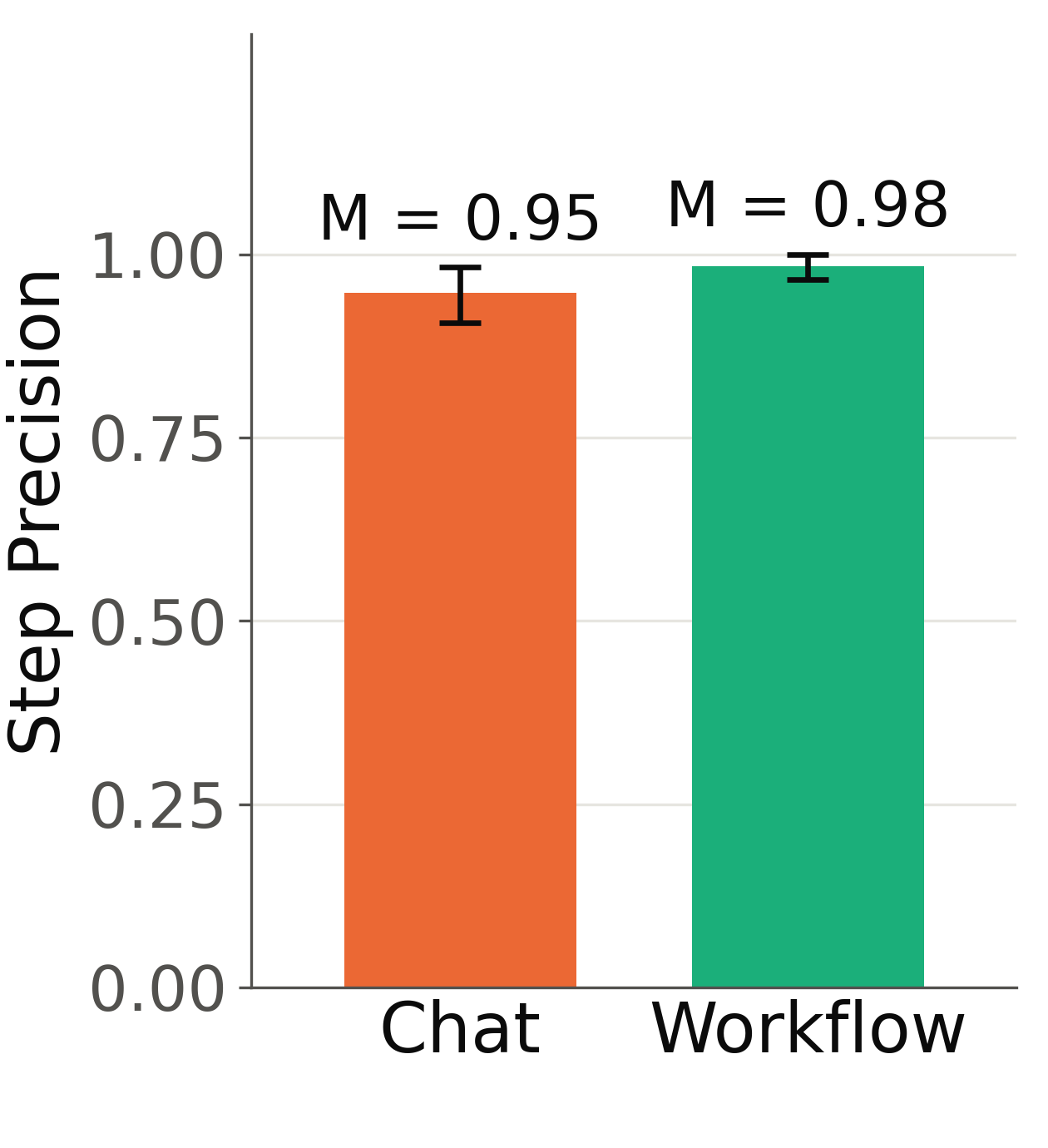}
        \caption{Step Precision}
        \label{fig:step_precision}
    \end{subfigure}
    \hfill
    \begin{subfigure}[t]{0.23\textwidth}
        \centering
    \captionsetup{skip=3pt}
    \includegraphics[
        width=\linewidth,
        trim=0 20 0 0,
        clip
    ]{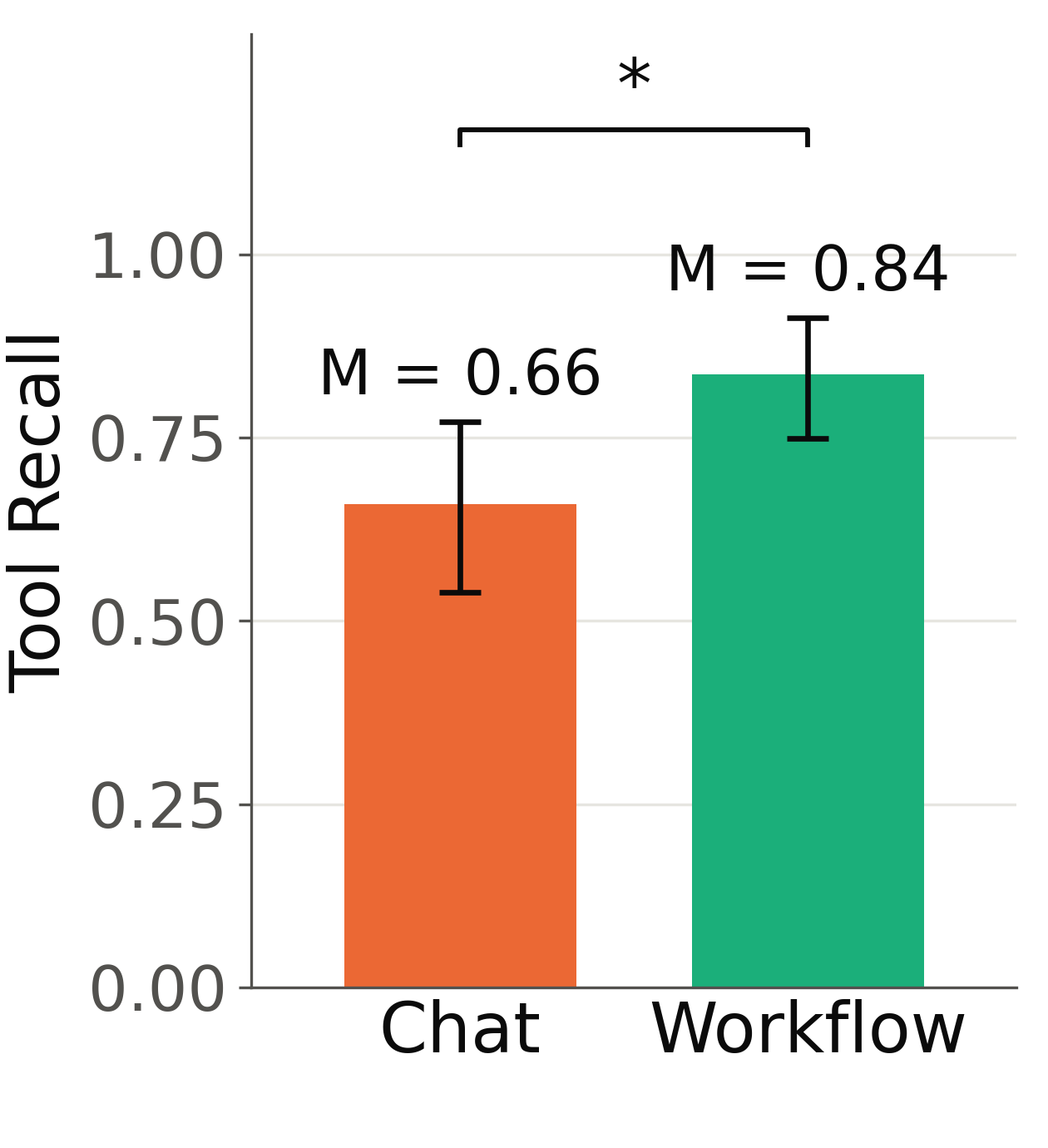}
        \caption{Tool Recall}
        \label{fig:tool_recall}
    \end{subfigure}
    \hfill
    \begin{subfigure}[t]{0.23\textwidth}
        \centering
    \captionsetup{skip=3pt}
    \includegraphics[
        width=\linewidth,
        trim=0 20 0 0,
        clip
    ]{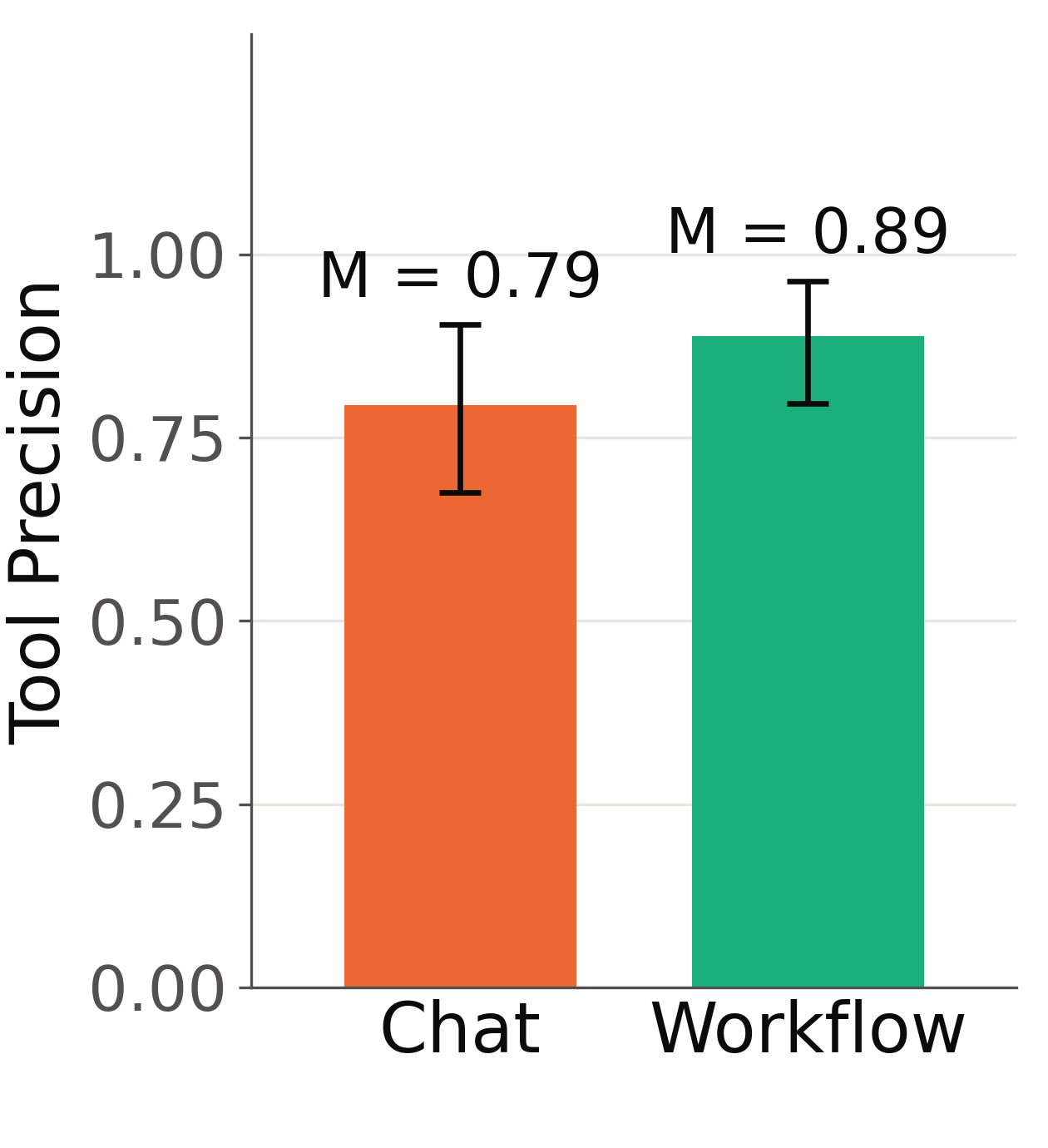}
        \caption{Tool Precision}
        \label{fig:tool_precision}
    \end{subfigure}

    \vspace{0.8em}

    \makebox[\textwidth][c]{%
        \begin{subfigure}[t]{0.23\textwidth}
            \centering
    \captionsetup{skip=3pt}
    \includegraphics[
        width=\linewidth,
        trim=0 20 0 0,
        clip
    ]{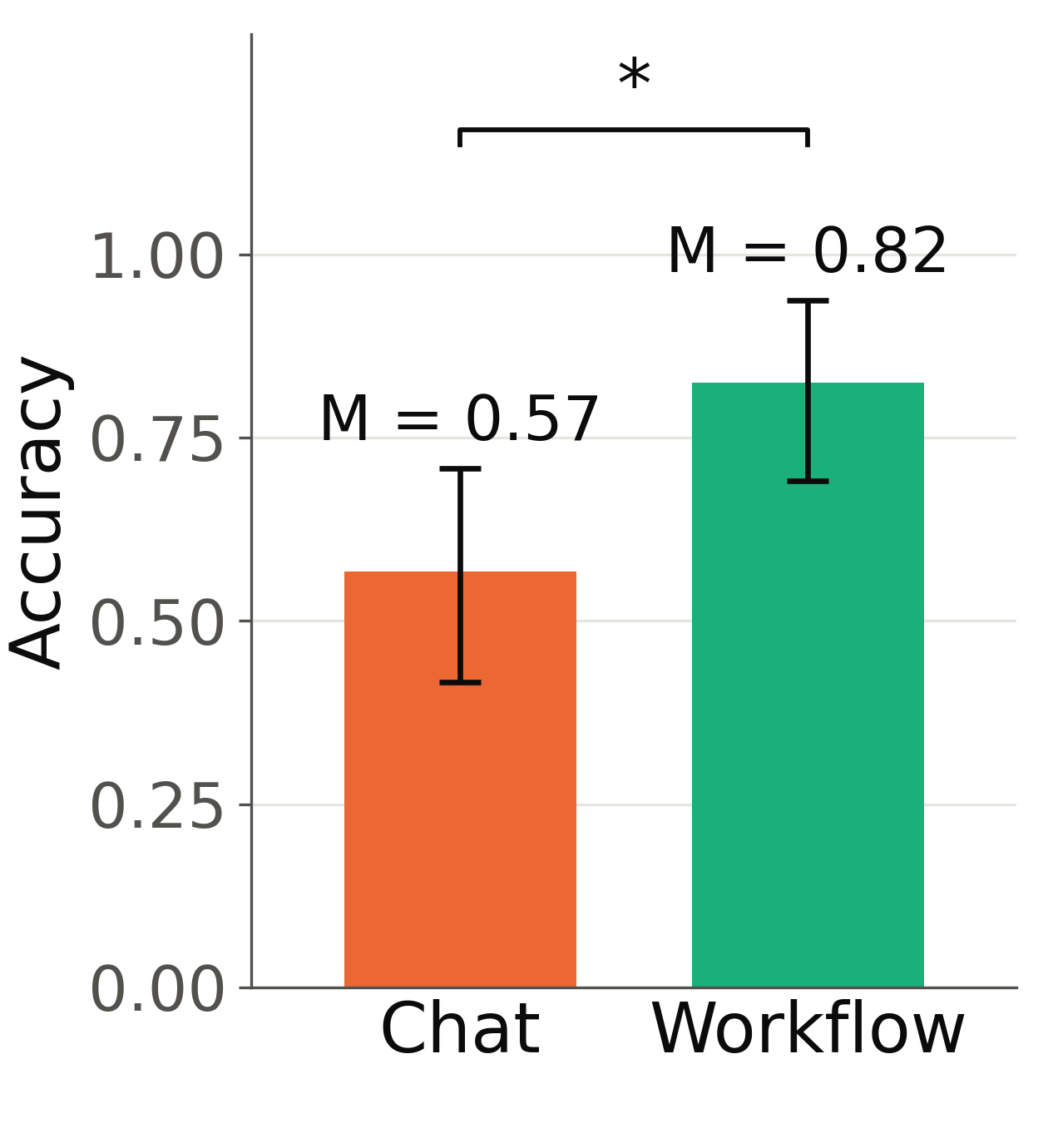}
            \caption{Process Comprehension}
            \label{fig:process_comprehension}
        \end{subfigure}
        \hspace{0.06\textwidth}
        \begin{subfigure}[t]{0.23\textwidth}
            \centering
    \captionsetup{skip=3pt}
    \includegraphics[
        width=\linewidth,
        trim=0 20 0 0,
        clip
    ]{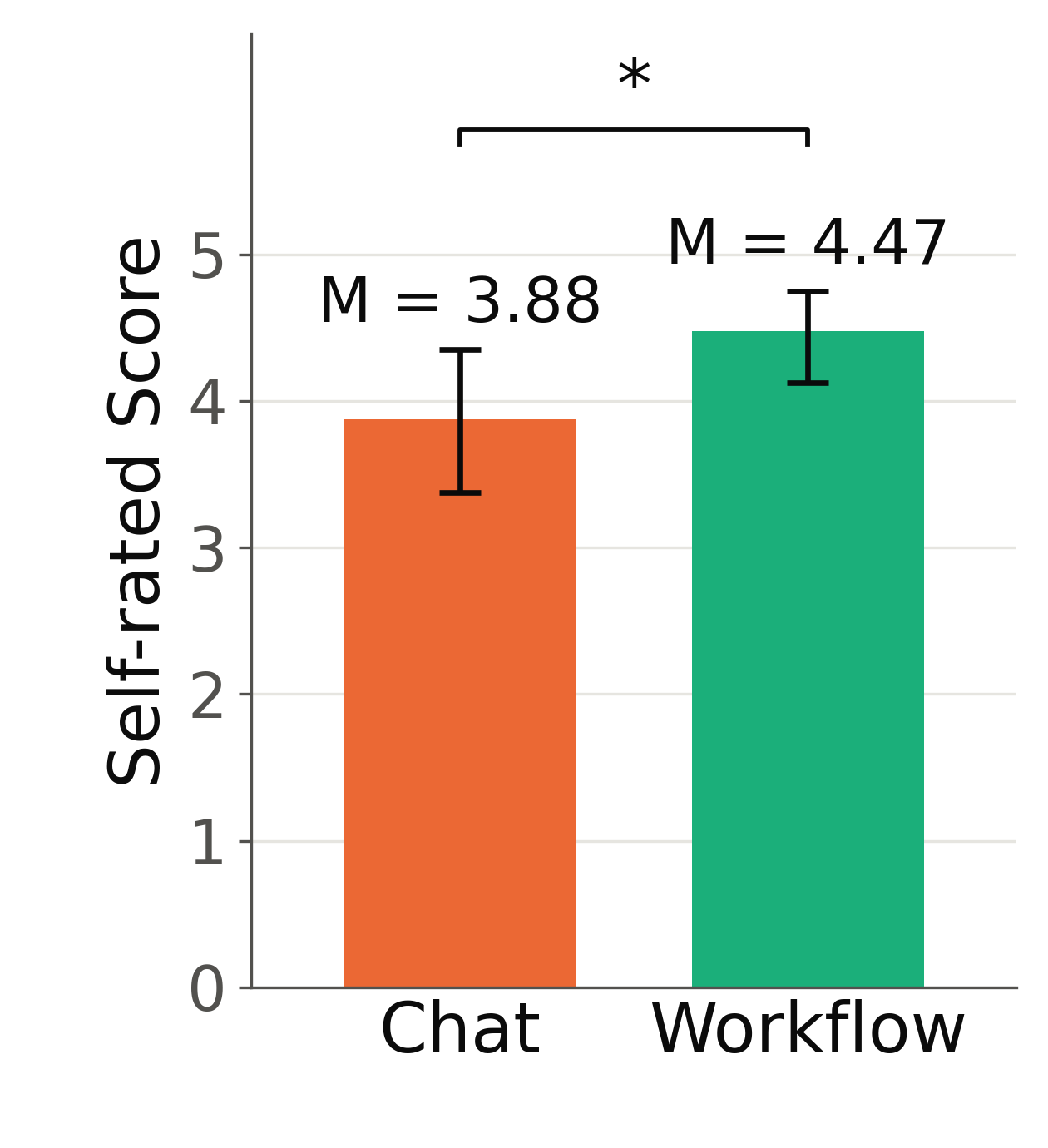}
            \caption{Self-reported Understanding}
            \label{fig:self_report_understanding}
        \end{subfigure}
    }

    \caption{Participants' understanding of the agent's execution in the chat and workflow conditions. Asterisks indicate significance: $^{*}p<.05$, $^{**}p<.01$.}
    \Description{Six comparisons of participants' understanding in the chat and workflow conditions. Step recall increases from 0.78 in chat to 0.94 with the workflow, tool recall from 0.66 to 0.84, process-comprehension accuracy from 0.57 to 0.82, and self-reported understanding from 3.88 to 4.47; these differences are statistically significant. Step precision is 0.95 versus 0.98 and tool precision is 0.79 versus 0.89, with no significant differences. Error bars show variability.}
    \label{fig:understanding_metrics}
\end{figure*}

We first examined participants’ understanding of the agent’s execution using step and tool precision and recall, process-comprehension accuracy, and self-reported understanding, as defined in \autoref{subsec:data_analysis}. As shown in \autoref{fig:understanding_metrics}, participants in the workflow condition recalled significantly more of the agent’s major steps ($M=.94$ vs. $.78$, $p<.01$) and tools ($M=.84$ vs. $.66$, $p<.05$). In contrast, precision was high in both conditions and did not differ significantly for either steps ($M=.98$ vs. $.95$) or tools ($M=.89$ vs. $.79$). Participants in the workflow condition also achieved significantly higher process-comprehension accuracy ($M=.82$ vs. $.57$, $p<.05$) and reported higher overall understanding ($M=4.47$ vs. $3.88$, $p<.05$).

\subsubsection{Validating the Agent’s Execution Performance: Outcomes and Patterns}

We next examined whether participants correctly judged task completion in the review task. Across task sessions, validation was successful in 15 of 23 workflow-condition sessions (65\%) and 7 of 23 chat-condition sessions (30\%). The logistic mixed-effects model showed a significant effect of condition, with higher validation success in the workflow condition ($\beta=2.62$, $SE=1.04$, $p<.05$). The model also showed a significant overall effect of task ($\chi^2(3)=8.66$, $p=.034$). Pairwise comparisons indicated that validation success was significantly higher for T3 than T4 ($OR=85$, $p=.019$), while session number was not significant.


\begin{table*}[t]
    \centering

    \begin{tabular}{p{0.04\textwidth} p{0.11\textwidth} p{0.43\textwidth} c c c}
        \toprule
        \textbf{TID} &
        \textbf{Error Type} &
        \textbf{Error Location} &
        \textbf{Chat} &
        \textbf{Workflow} &
        \textbf{$\Delta$} \\
        \midrule

        T1
        & Agent error
        & The CV contains a fictional location, but the agent treated it as a valid European location.
        & 2/6 (33\%)
        & 3/6 (50\%)
        & +17\% \\

        \midrule
        T2
        & Prompt error
        & The prompt mixed up which article group should be sent to Slack and which should be written to Google Sheets.
        & 1/6 (17\%)
        & 5/6 (83\%)
        & +66 \%  \\

        \midrule
        T3
        & Prompt error
        & The prompt omitted the requirement to write non-fiction books to Sheet 3.
        & 4/6 (67\%)
        & 5/5 (100\%)
        & +33 \%  \\

        \midrule
        T4
        & Agent error
        & The agent read the existing reviews but failed to use them to filter out old reviews before updating Google Sheets.
        & 0/5 (0\%)
        & 2/6 (33\%)
        & +33 \%  \\

        \bottomrule
    \end{tabular}

    \caption{Validation accuracy across review tasks and conditions. 
    $\Delta$ indicates the percentage-point difference between the workflow and chat conditions.}
    \label{tab:validation_by_task}
\end{table*}

As shown in \autoref{tab:validation_by_task}, the workflow condition had a higher validation success rate than the chat condition for each task, although the size of this descriptive difference varied across tasks. 

The smallest difference appeared in T1 (33\% vs. 50\%), where the error came from the agent incorrectly treating a fictional location as European. Detecting this error required participants to question a specific piece of content in the CV, and the workflow provided little additional process-level evidence for making this judgment.

The workflow provided clearer support in T2 and T3, where the errors were directly reflected in the workflow structure. In T2, the relevant and non-relevant article groups were routed to the wrong outputs, producing a large difference between the chat and workflow conditions (17\% vs. 83\%). In T3, the task required non-fiction books to be written to Sheet 3, but the corresponding workflow branch was missing; validation success was already relatively high in the chat condition, but increased from 67\% to 100\% with the workflow. 


T4 was the most difficult task in both conditions. No participant in the chat condition correctly identified the error, compared with 2 of 6 participants in the workflow condition. Here, the agent read the existing reviews but did not use them to determine which reviews were new before writing to Google Sheets. This problem was difficult to detect from the final output alone, while the workflow exposed the missing dependency between these steps. However, this cue was subtle, and most participants in the workflow condition still overlooked it.


\begin{figure*}[t]
    \centering

\begin{minipage}[t]{0.25\textwidth}
    \vspace{0pt}
    \centering
    \captionsetup{skip=3pt}
    \includegraphics[
        width=\linewidth,
        trim=0 20 0 0,
        clip
    ]{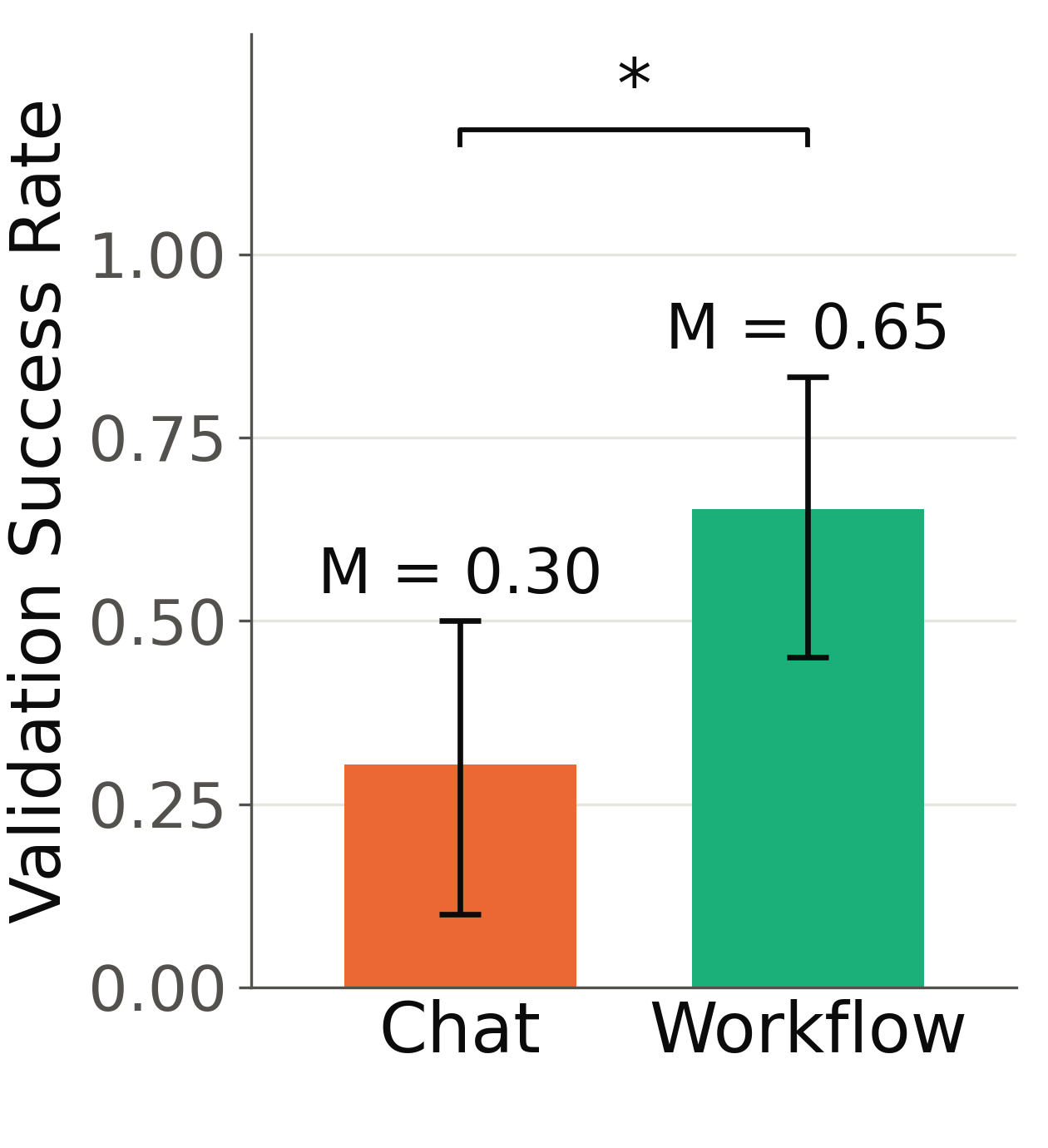}
    \captionof{figure}{Validation success rate across conditions.}
    \Description{Bar chart comparing validation success across conditions. Participants correctly validated 30 percent of executions in the chat condition and 65 percent in the workflow condition, showing significantly higher validation success when the post-task workflow was available. Error bars show variability.}
    \label{fig:validation_success}
\end{minipage}
    \hfill
    \begin{minipage}[t]{0.72\textwidth}
        \vspace{30pt}
        \centering
        \small
        \begin{tabular}{@{}lccc@{}}
            \toprule
            \textbf{Review Pattern} &
            \textbf{Chat} &
            \textbf{Workflow} &
            \textbf{Overall} \\
            \midrule

            No pattern observed
                & 0/3 (0\%) & -- & 0/3 (0\%) \\

            Output check only
                & 1/10 (10\%) & -- & 1/10 (10\%) \\

            Workflow check only
                & N/A & 0/2 (0\%) & 0/2 (0\%) \\

            Output + workflow check
                & N/A & 4/8 (50\%) & 4/8 (50\%) \\

            Output (+ workflow) + cross-check
                & 6/10 (60\%) & 11/13 (85\%) & 17/23 (74\%) \\

            \bottomrule
        \end{tabular}

\captionof{table}{Review patterns observed when participants assessed the AI agent's task execution. Each cell reports the number of successful validations / number of sessions in which the pattern was observed, with the corresponding success rate in parentheses. ``--'' indicates that the pattern was not observed in that condition; N/A indicates that the pattern was not applicable because the workflow was unavailable.}
        \label{tab:review_patterns}
    \end{minipage}

\end{figure*}

To better understand how participants reached these judgments, we examined their review actions across all task sessions. We identified three common types of review actions:  \textit{output check},\textit{workflow check} and  \textit{cross-check}. \textit{Output check} refers to inspecting the agent’s final outputs, such as rows written to Google Sheets or messages posted to Slack. \textit{Workflow check} refers to inspecting the workflow structure or individual nodes, including their inputs and outputs. \textit{Cross-check} refers to comparing information across different sources, such as checking the task prompt against the agent’s output, or rerunning a workflow node and comparing its result with the final output. Participants often combined these actions within the same session, leading to the review patterns summarized in Table~\ref{tab:review_patterns}.

The results in Table~\ref{tab:review_patterns} show that validation was more successful when participants combined multiple sources of evidence. Sessions that relied only on output checking had a low success rate (1/10, 10\%), and neither of the two sessions that relied only on workflow checking resulted in a correct judgment. In contrast, participants succeeded in 4 of 8 sessions (50\%) when they inspected both the final output and the workflow, but used them as separate sources of information. When participants went a step further and cross-checked information across sources, the success rate increased to 17 of 23 sessions (74\%) overall. This pattern was especially clear in the workflow condition, where participants who combined output checking, workflow checking, and cross-checking correctly validated the execution in 11 of 13 sessions (85\%).


\subsubsection{Interview Results}

Participants’ interview responses further revealed that the value of the post-task workflow was not simply in showing more information, but in changing how participants reviewed the agent’s work. Seventeen of the 20 participants (P1,P2,P4,P5,P7-P14,P16-P20) found the post-task workflow helpful when reviewing the agent’s work. Two saw little added value (P3,P6), and one found the graph more confusing than helpful (P15). The main benefit, mentioned by 14 participants (P5,P7-P14,P16-P20), was that the workflow made the agent’s process transparent. P20 described prompt-based interaction as “\textit{opaque},” while P18 said that it mainly showed “\textit{the final output or maybe some logs}.” By contrast, P13 said that the workflow exposed “\textit{the step-by-step structure and the control flow},” and P17 emphasized that it showed “\textit{how the task is being processed}.” Seven participants also valued being able to open individual steps and examine their inputs, outputs, parameters, and tool-call results (P5,P7,P8,P9,P16,P18,P20). P8 noted that this provided access to “\textit{the responses to tool calls, not just the final response}.” P11 explained, this step-level view was especially useful when something looked wrong as it allowed them to see “\textit{at which step things go wrong}.”

The interviews also shed light on how participants arrived at their judgments of the agent’s performance, including why some assessments were unsuccessful. When participants relied mainly on the final output, they often checked whether the expected files, entries, or other outputs were present rather than whether their contents were correct. For example, P1, who incorrectly judged T1 as successful, explained: “\textit{when I first opened the folder there were three CVs, and in the final sheet there were also three entries, and one of them said ‘fit’—so I assumed it worked}.” P1 further described a tendency to accept the output unless something clearly appeared wrong: “\textit{I didn’t debug it line by line … it seems to comply with the requirements—like when you get your bank statement, you assume the summary is correct without checking each entry.}” This became especially difficult when tasks produced many results. P3 noted, “\textit{I didn’t check the final sheet … because it already has a lot of entries—it’s long},” while P6 similarly said, “\textit{I didn’t read every comment individually, so I’m not sure}.”

The interviews provided more direct evidence of how the workflow could support validation when an error was reflected in the execution structure. Among participants who successfully identified the errors in these two tasks, 70\% explicitly referred to the workflow as helping them notice the problem. In T3, for example, the omitted operation of writing results to a third sheet appeared as a missing branch in the workflow. P10 explicitly used this structural absence as evidence, observing that “\textit{with the two branches we are updating the first sheet and the second sheet … but we are missing the third sheet. There should be a third branch}.” Similarly, P6 noted that “\textit{it logs everything it’s doing, and there is no ‘writing to Sheet 3’ step}.”The Appendix provides visual examples showing how each selected error appeared in the corresponding task.

\subsection{Completing Follow-up Tasks: Workflow Reuse vs. Prompting}

\subsubsection{Follow-up Task Outcomes: Accuracy, Efficiency, and Perceived Difficulty}

\begin{figure*}[t]
    \centering

    \begin{subfigure}[t]{0.31\textwidth}
        \centering
        \captionsetup{skip=3pt}
        \includegraphics[
            width=\linewidth,
            trim=0 20 0 0,
            clip
        ]{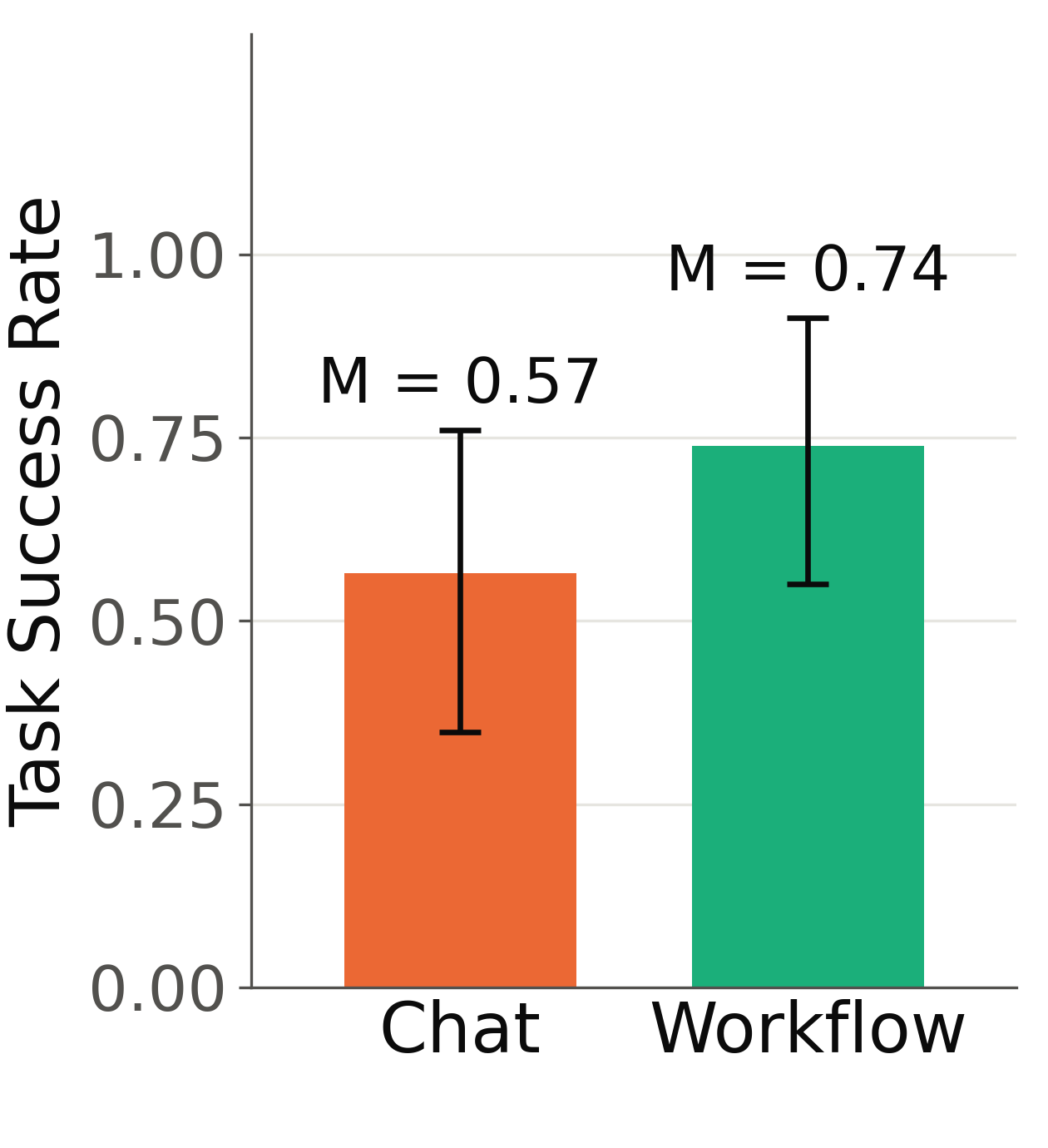}
        \caption{Task Success Rate}
        \label{fig:followup_success}
    \end{subfigure}
    \hfill
    \begin{subfigure}[t]{0.31\textwidth}
        \centering
        \captionsetup{skip=3pt}
        \includegraphics[
            width=\linewidth,
            trim=0 20 0 0,
            clip
        ]{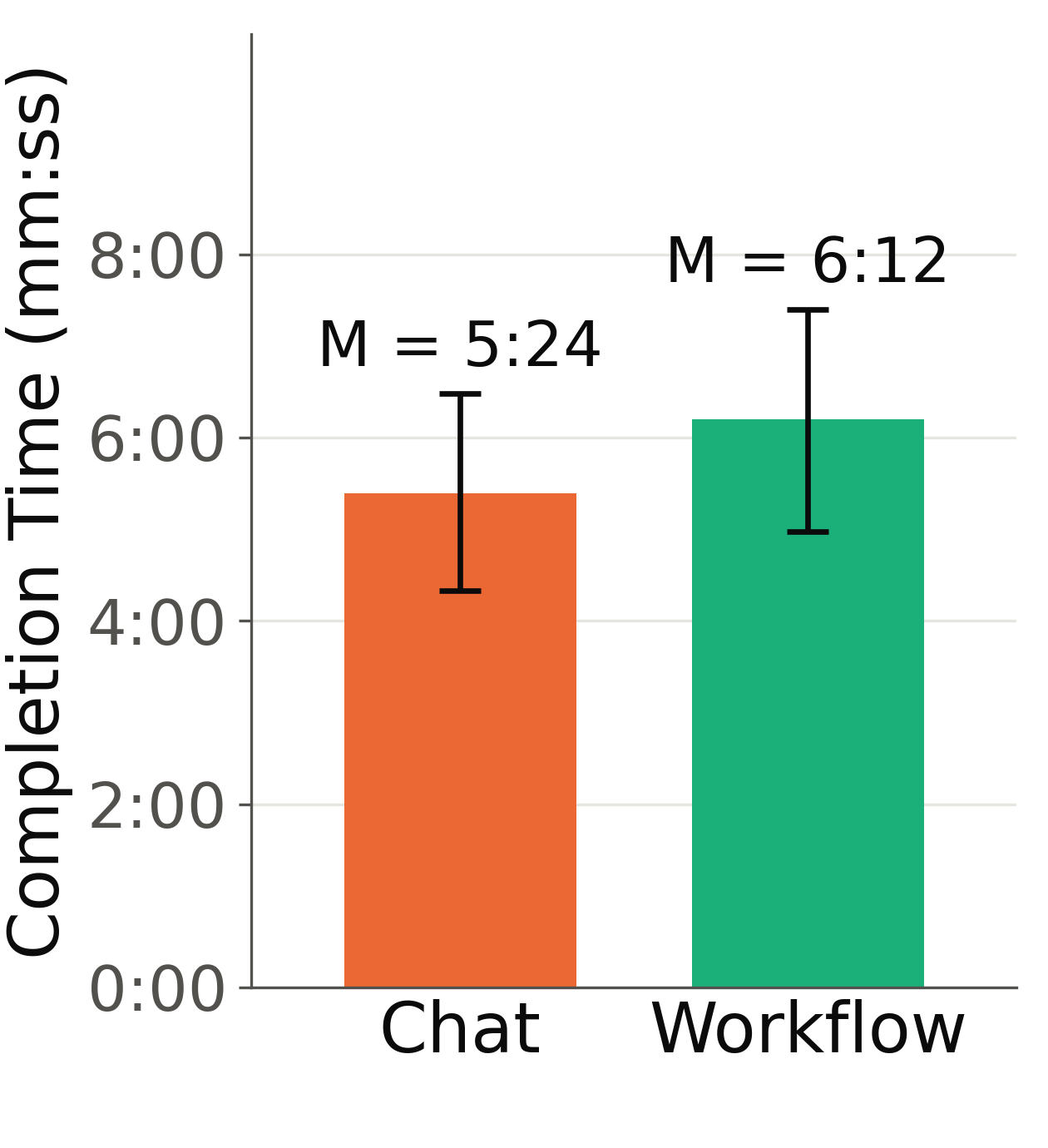}
        \caption{Completion Time}
        \label{fig:followup_time}
    \end{subfigure}
    \hfill
    \begin{subfigure}[t]{0.31\textwidth}
        \centering
        \captionsetup{skip=3pt}
        \includegraphics[
            width=\linewidth,
            trim=0 20 0 0,
            clip
        ]{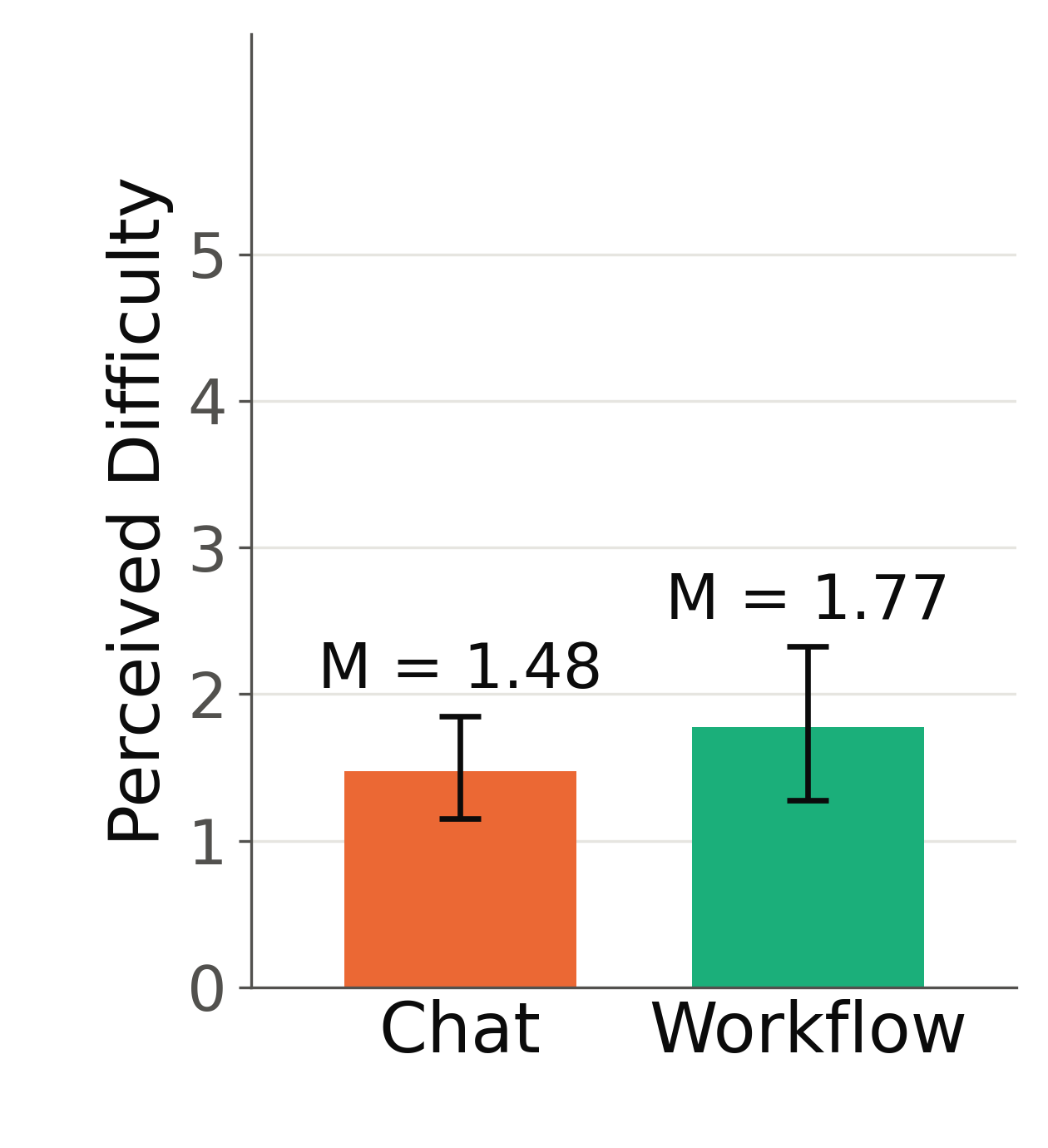}
        \caption{Perceived Difficulty}
        \label{fig:followup_difficulty}
    \end{subfigure}

    \caption{Performance in the follow-up task across the chat and workflow conditions. Differences were not statistically significant. }

    \Description{Three comparisons of follow-up-task performance in the chat and workflow conditions. Task success is 0.57 in chat and 0.74 with the workflow; mean completion time is 5 minutes 24 seconds in chat and 6 minutes 12 seconds with the workflow; and perceived difficulty is 1.48 versus 1.77. None of these differences is statistically significant. Error bars show variability.}
    \label{fig:followup_performance}
\end{figure*}

We examined follow-up-task performance in terms of task success, completion time, and perceived difficulty. As shown in \autoref{fig:followup_success}, across task sessions, task success was 17 of 23 in the workflow-condition sessions (74\%) and 13 of 23 chat-condition sessions (57\%). The logistic mixed-effects model found no significant effect of condition on task success ($p\geq.05$); neither task nor session number was significant. 

For completion time and perceived difficulty (as shown in \autoref{fig:followup_time} and \autoref{fig:followup_difficulty}), the participant-level comparisons showed no significant differences between conditions. Participants took bit longer to complete the task in the workflow condition than in the chat condition ($M=6$ min $12$ sec vs. $5$ min $24$ sec), while perceived difficulty was similar between conditions ($M=1.77$ vs. $1.48$).



\subsubsection{Interview Results}

The interviews revealed a more divided picture of workflow reuse than of workflow-based review. Of the 20 participants, nine preferred adapting the existing workflow for the follow-up task (P1, P2, P4, P9, P10, P11, P14, P16, P18), six preferred editing the previous prompt (P3, P6, P7, P13, P15, P19), and five said that their preference depended on the situation (P5, P8, P12, P17, P20).

Among participants who favored the workflow, a key advantage was that it made changes easier to locate and carry out. Five participants (P1, P9, P10, P11, P16) said that the graph helped them quickly find the part of the task that needed to change, without rereading and interpreting the whole prompt. P1 explained, “\textit{when you copy-paste the old prompt, it's harder to find, among all the text, the exact place to make the change. With the visual workflow it's easier to spot.}” Once the relevant part was located, three participants (P10, P14, P16) also found the workflow easier to edit because the task was already broken into explicit steps. This was particularly useful for local structural changes, such as adding or removing an operation. As P14 explained, “\textit{some things—like adding or removing nodes in similar tasks—are yes/no choices that don't need to be expressed through language at all.}”

The workflow also gave participants ways to check their changes before committing to the full task. Six participants (P5, P8, P9, P11, P18, P20) valued being able to inspect and test individual steps during adaptation, echoing the usefulness of step-level inspection that also emerged in the review task. Here, however, the benefit was different: participants could check whether a modified step behaved as expected before rerunning the entire workflow. This sense of working with an existing, adaptable process also made the workflow feel more concrete and controllable. Five participants (P8, P9, P10, P14, P18) also described the existing workflow as making the adaptation process feel more concrete and controllable. P18 compared it to a “\textit{toolbox}” that “\textit{keeps everything in control,}” while describing modification through the previous prompt as more “\textit{fragile.}”

Participants who preferred prompting pointed to a different set of costs. Five participants  (P3,P6,P7,P13,P15) highlighted the learning required to understand and edit the workflow. P7 explained, ``\textit{I'm not that familiar with what each module does, so I hesitate to delete or modify things.}'' For these participants, changing the prompt felt more familiar and required less knowledge of the workflow representation. As P15 put it, ``\textit{to add nodes or change things I need to understand more; with the prompt it was quick.}'' Three participants (P3,P5,P20) also noted that workflow editing could become tedious when the required changes were numerous, spread across different parts of the graph, or involved a large workflow. Importantly, this preference for prompting largely depended on having the previous prompt available for adaptation. When we asked the six participants who preferred prompting whether they would still choose this approach if they had to write a new prompt from scratch, only P15 said she would continue to prefer prompting. This suggests that, for most of these participants, the appeal of prompting was tied to prompt reuse rather than prompting itself. In practice, however, prior prompts may not always be readily available or easy to retrieve, which could make workflow reuse more attractive as a persistent artifact for related future tasks.

For the five participants whose preference depended on the situation, the clearest dividing line was the scope of the adaptation. Small and localized changes favored the workflow, while broader changes that affected much of the task favored prompting. P20 summarized this distinction: ``\textit{if I only need to add one operation or one new branch---small, local modifications---the workflow is fine. But if I'd have to substantially modify many of the task's nodes, including restructuring, then I think the prompt is more suitable. So it mostly depends on the degree of adaptation.}''

Beyond these reasons, participants' preferences appeared to be associated with their prior experience with workflow-automation tools. Among the seven participants with recent workflow-automation experience (\autoref{subsubsec:recruitment}), six preferred adapting the workflow for the follow-up task (P2, P4, P9, P10, P14, P18), while one said that their preference depended on the situation (P5).Among the 13 participants without such experience, preferences were more mixed: three preferred the workflow (P1, P11, P16), six preferred prompting (P3, P6, P7, P13, P15, P19), and four said that their preference depended on the situation (P8, P12, P17, P20). The reasons participants gave point to a likely explanation: all five participants who cited the learning required to understand and edit the workflow came from the group without workflow experience. Given the small subgroups, we treat this as an exploratory observation rather than a tested effect, but it suggests that familiarity with workflow representations lowers the threshold for reusing them, whereas participants encountering such representations for the first time more often fell back on the prompt. 


\section{Discussion and Conclusion}

\subsection{
Post-Task Workflows as Post Hoc Procedural Transparency}

Our findings return to the process-transparency challenge discussed in \autoref{subsec:process_transparency}: how to help users understand an agent’s execution without requiring continuous monitoring. Consistent with DG1, the post-task workflow addressed this challenge by externalizing the completed execution as a structured process. Although, participants could see the original prompt, execution progress, and final output in the chat condition, these pieces did not necessarily provide a coherent picture of how the agent carried out the task as a whole. With the workflow, participants could see the major steps, tools, dependencies, and overall execution structure in one representation. This was reflected in our quantitative results, where participants in the workflow condition recalled more of the agent's major steps and tools and showed higher process comprehension. 


The value of a post-task workflow therefore appears to lie not simply in exposing more execution information, but in organizing information distributed across the prompt, execution trace, tool activities, and final output into a representation of how these pieces relate. Participants similarly described the workflow as making the agent's process more transparent by exposing its ``\textit{step-by-step structure and the control flow.}'' In this sense, post-task workflows complement existing forms of agent transparency: lightweight execution updates can indicate what the agent is currently doing, while the workflow provides a persistent overview after execution has finished. This may be particularly useful for agent tasks in which users do not continuously monitor execution but still need to understand afterwards how the result was produced.

Beyond supporting understanding, this process transparency also changed how participants validated the agent’s execution. With the workflow available, participants could move between the prompt, final output, workflow structure, and individual nodes rather than judging the task mainly from its final output. This difference was also visible in their review patterns (Table~4): none of the workflow-condition sessions relied on output-checking only or showed no observed review pattern; participants always engaged in workflow checking, cross-checking, or both. By contrast, output-checking only was relatively common in the chat condition and was associated with low validation success. Our observed review patterns further suggest that cross-checking was important: workflow checking alone was not sufficient for successful validation, but participants were substantially more successful when they combined output checking, workflow checking, and cross-checking across different sources of evidence. The post-task workflow therefore appears to function less as a replacement for existing information and more as an additional reference point that connects these different forms of evidence. This suggests that making the agent’s execution process visible and traceable can encourage users to take a more active role in validating the execution, rather than relying primarily on the final output when judging whether the task was completed correctly.

Our findings connect to a broader concern about the human oversight of AI systems. Oversight of agentic systems differs from oversight of generative systems in that users must assess not only what was produced but how it was produced, and current agent interfaces support this poorly because the record of an execution is long, linear, and dense~\cite{passi2025agentic}. Our chat condition reflects what agent interfaces provide today, and participants judged the execution correctly in fewer than a third of sessions and described difficulties in inspecting long outputs. The post-task workflow gives users a structure against which parts of an execution can be checked selectively rather than exhaustively, providing some insights into how a system arrived at its result, important for enabling effective human oversight~\cite{faas2026, passi2025agentic, sterz2024}.

\subsection{What Workflows Reveal—and What They Do Not}
In terms of validation and identifying errors, the workflow was especially useful when the error was reflected in the execution structure. This was evident in the task-level results in \autoref{tab:validation_by_task}: in T2 and T3, participants used the workflow to identify incorrect routing or a missing branch. T4 provided an even stronger example. Although the final output did not reveal that the agent had failed to use the existing reviews when determining which reviews were new, the workflow exposed the missing dependency between these steps. This highlights an important distinction between validating the outcome and validating the execution process: a correct-looking output does not necessarily mean that the task was carried out correctly or will be in future runs. This extends process transparency from supporting a general understanding of \emph{what the agent did} toward supporting targeted investigation of \emph{where and how the process went wrong}.

Meanwhile, our findings also indicate a boundary of workflow-based transparency. Even with the workflow available, participants reached an incorrect judgment in 8 of 23 sessions (35\%). The workflow provided less benefit when an error depended mainly on judging the correctness of specific content rather than the structure of the execution. For example, in the review task of T1,  when identifying an error required recognizing that a particular location in a CV was fictional, the workflow offered little additional evidence beyond the underlying content itself. Post-task workflows therefore are not sufficient on their own for validating every aspect of an agent’s execution. When an error depends on correctly interpreting or verifying the underlying task data, users may still need to inspect the source inputs, intermediate results, or final outputs directly. Workflow-based review should therefore link each step to the exact data and artifacts it consumed and produced, so that users can verify not only how the execution proceeded, but also whether the information underlying each step was correct.


\subsection{
Post-Task Workflows as a Persistent Artifact}

When users need to carry a completed agent process forward to a related task, our findings suggest that post-task workflows can provide an alternative to prompting. In the follow-up task, we found no significant differences between workflow and prompt reuse in task success, completion time, or perceived difficulty. Rather than showing that one approach was consistently better, these results show that an existing workflow can be adapted to complete a related task alongside adapting a previous prompt.

Our interviews further show that participants’ preferences between these two reusable artifacts depended on their prior workflow-automation experience, what was available to reuse, and the scope of the required adaptation. Prior workflow-automation experience appeared to lower the barrier to workflow reuse, while participants without such experience more often found prompting familiar and easier to modify. The scope of adaptation also mattered: localized changes were easier to map onto particular workflow steps, whereas broader changes requiring substantial restructuring favored prompting. Importantly, however, participants’ preference for prompting largely depended on having the previous prompt available. Of the six participants who initially preferred prompt reuse, only P15 still preferred prompting when asked to imagine writing the prompt from scratch. This suggests that the practical value of prompt reuse depends heavily on whether a prior prompt is actually available, whereas a post-task workflow preserves the executed process as a reusable artifact.

\subsection{Limitations and Future Works}
\label{sec:limitations}
Our study has several limitations that suggest directions for future work. First, we studied a relatively small sample of experienced AI-agent users, with only a smaller subset (N=7) having recent workflow-automation experience. Our observation that prior workflow experience may lower the barrier to workflow reuse should therefore be treated as exploratory. Future studies with larger and more diverse samples could examine this relationship more systematically. Longer-term studies are also needed to understand whether the learning cost experienced by workflow novices persists or decreases as users become familiar with workflow representations. Preferences might also change if users have personal stakes in the automated task or if auditing or human oversight are demanded by legal or company regulations~\cite{LamAudit2024, sterz2024, faas2026}. 

Second, our study focused on a compact set of tasks with limited size and structural complexity, and the follow-up tasks were designed to preserve the structure of the original task with little modification. This allowed us to study workflow adaptation and reuse under controlled conditions, but limits what we can conclude about larger workflows or follow-up tasks that require broader changes to the existing process. Future work could systematically vary the scope of adaptation to understand when users benefit from adapting a post-task workflow and when prompting becomes more appropriate. Similarly, our validation findings were based on a limited set of prompt errors and agent errors. Examining a broader range of execution errors would help clarify which failures are reflected in the execution structure and which require users to check the underlying data, outputs, or domain-specific evidence.

Finally, Trace2Flow is intended as a research probe for studying the role of post-task workflows for end users, rather than as a general method for trace-to-workflow translation studied in prior work~\cite{liu2026flowmind, liu2026reuseit}. We provide an additional validation of the translation approach on the study tasks in the Appendix, but this evaluation is limited to the task set used in our study and does not establish how well the approach generalizes across broader tasks, tools, or AI agents. 
Future work could evaluate trace-to-workflow translation across larger and more diverse task sets, and examine how different translation approaches affect users’ understanding, validation, and reuse of the resulting post-task workflow.

\bibliographystyle{ACM-Reference-Format}
\bibliography{sample-base}

\appendix

\end{document}